# Processing of synthetic data in AI development for healthcare and the definition of personal data in EU law

**Authors:**

Vibeke Binz Vallevik,

> Title and job description: MSc, Senior Principal Researcher AI in Healthcare,
> Affiliations: DNV and University of Oslo
> Address: Veritasveien 1, 1322 Høvik, Norway
> Email: vibeke.binz.vallevik@dnv.com

Anne Kjersti C. Befring,

> Title and job description: Professor of Law Affiliation: University of Oslo, Department of Public and International Law
> Address: PB 6706 St Olavs plass, NO-0130 Oslo, Norway
> Email: a.k.befring@jus.uio.no; https://www.jus.uio.no/ior/personer/vit/akbefrin/

Severin Elvatun,

> Title and job description: PhD, Data scientist
> Affiliation: The Cancer Registry of Norway, Norwegian Institute of Public Health
> Address: Ullernchausseen 64, 0379 Oslo, Norway
> Email: sela@kreftregisteret.no

Jan Franz Nygård,

> Title and job description: PhD, Head of the Registry Informatics Department
> Affiliations: The Cancer Registry of Norway, Norwegian Institute of Public Health
> and UiT- The Arctic University of Norway. Address: Ullernchausseen 64, 0379 Oslo, Norway
> Email: jfn@kreftregisteret.no

Processing of synthetic data in AI development for healthcare and the definition of personal data in EU law

# Contents

# Processing of synthetic data in AI development for healthcare and the definition of personal data in EU law

## Abstract

Artificial intelligence (AI) has the potential to transform healthcare, but it requires access to health data. Synthetic data that is generated through machine learning models trained on real data, offers a way to share data while preserving privacy. However, uncertainties in the practical application of the General Data Protection Regulation (GDPR) create an administrative burden, limiting the benefits of synthetic data.

Through a systematic analysis of relevant legal sources and an empirical study, this article explores whether synthetic data should be classified as personal data under the GDPR. The study investigates the residual identification risk through generating synthetic data and simulating inference attacks, challenging common perceptions of technical identification risk. The findings suggest synthetic data is likely anonymous, depending on certain factors, but highlights uncertainties about what constitutes 'reasonably likely' risk. To promote innovation, the study calls for clearer regulations to balance privacy protection with the advancement of AI in healthcare.

# Processing of synthetic data in AI development for healthcare and the definition of personal data in EU law

## 1 Introduction

Artificial intelligence (AI) and data-driven technologies hold immense potential to transform healthcare by improving efficiency and effectiveness. The European Commission actively promotes the accessibility, usability and exchange of healthcare data to foster AI development.[1,2] However, despite these efforts, varying interpretations of the EU General Data Protection Regulation (GDPR) have complicated the sharing of health data since its implementation in 2018, thereby hindering AI-driven research and treatment development in healthcare.[3,4] There is a high cost of data collection and, particularly in rare diseases, a scarcity of patients available for inclusions in studies. Patient confidentiality is a prerequisite for keeping the population's trust in the healthcare system and consequently data processing is highly restricted. The need to access data to develop new tools and knowledge is at odds with this reality, and concerns regarding the limited amount of real-world data has sparked the interest in synthetic data generation.[5] Generation of synthetic data that mimics the statistical properties of a real dataset is proposed as a solution to improve access, efficiency and cost reduction in cases of data scarcity or high privacy risks.[6,7,8,9] Despite being referred to as an anonymization technique in The Data Governance Act (DGA), literature warns about a residual

---

[1] European Commission: Directorate-General for Communications Networks Content and Technology and PwC, *Study on eHealth, interoperability of health data and artificial intelligence for health and care in the European Union – Final study report. Lot 2, Artificial Intelligence for health and care in the EU* (2021).

[2] European Commission, *A European strategy for data COM/2020/66 final (EU data strategy)* (2020)

[3] Regulation (EU) 2016/679 of the European Parliament and of the Council of 27 April 2016 on the protection of natural persons with regard to the processing of personal data and on the free movement of such data (The General Data Protection Regulation)' (GDPR)

[4] Cristina Legido-Quigley et al., 'Data sharing restrictions are hampering precision health in the European Union' 31 Nature Medicine 360, 2025.; Jakov Vukovic et al., 'Enablers and barriers to the secondary use of health data in Europe: general data protection regulation perspective' 80 Archives of Public Health 115, 2022.

[5] Pablo Villalobos et al., 'Position: Will we run out of data? Limits of LLM scaling based on human-generated data' (Forty-first International Conference on Machine Learning, 2024)

[6] Steven M Bellovin, Preetam K Dutta and Nathan Reitinger, 'Privacy and synthetic datasets' 22 Stan Tech L Rev 1, 2019.

[7] Synthetic data is "data which is artificially generated as opposed to data captured in the real world", as defined by Clara Alloza et al., 'A Case for Synthetic Data in Regulatory Decision-Making in Europe' 114 Clinical Pharmacology & Therapeutics 795, 2023.

[8] The European Parliament actively supports synthetic data development for AI algorithm training to develop new treatment methods while protecting personal data and patient privacy, see: European Parliament, *European Parliament resolution of 18 December 2019 on enabling the digital transformation of health and care in the Digital Single Market; empowering citizens and building a healthier society (2019/2804(RSP))* (Document 52019IP0105 edn, Eurlex, 2021).

[9] An example of synthetic data sharing from the financial sector: the EU's digital finance hub where synthetic data is used to facilitate low-risk data sharing, see 'European Commission, The Digital Finance Platform- Synthetic Data' <https://digital-finance-platform.ec.europa.eu/data-hub/what-kind-of-data-available> accessed 19.11.2024.

Processing of synthetic data in AI development for healthcare and the definition of personal data in EU law identification risk.[10,11] This leads to a question of whether synthetic data should be considered personal data, which in turn creates hesitation in its use and limits its full potential.[12]

The GDPR regulates processing of personal data within the EU, setting standards for data collection, sharing and protection. The EU AI Act is harmonized with GDPR, stating that the GDPR applies to the processing of personal data in AI models.[13]

This article examines whether synthetic data falls under the legal definition of personal data in the GDPR art. 4 (1) and consequently, whether it is subject to the regulation's provisions under EU law. By exploring this legal classification, the article aims to clarify the conditions under which synthetic data can be processed and shared within the boundaries of EU law. If synthetic data is not considered personal data, it will fall outside the scope of the GDPR but may still be covered by the AI Act which regulates the risks associated with the use of AI and does not distinguish between data that can be linked to individuals or not. While personal data in healthcare is assumed to be 'data concerning health' under GDPR articles 4(15) and 9, the article assumes there has been a legal basis for processing the healthcare data for the synthetization itself and focuses solely on the legal definition of the synthesized dataset. Synthetic datasets that are considered to be personal data will be subject to the restrictions and security measures that follow from the GDPR and must be processed lawfully, fairly and transparently and in accordance with the data protection principles of GDPR art. 5. In addition, a basis for processing under GDPR art. 6 and an exception to the prohibition on processing health data under GDPR art. 9 are required.

The thesis of this article is that in evaluating identifiability according to GDPR art. 4(1), the mere existence of additional information that may aid in identification of synthetic datasets is often overemphasized at the expense of assessing the actual practical identification risk.

The article first outlines the methods and legal sources relevant for the analysis, focusing on the GDPR and the AI Act and how EU primary law and secondary law come into play. The focus in this article is on synthetic data that has been generated by training machine learning models on a real health dataset to create a synthetic version that does not have a one-to-one relationship with the original datapoints. The actual risk of identification and proportionality is a basis for the legal assessments. An empirical study has therefore been conducted to map the risk in this type of setting and to provide the factual background for synthetic data generation. Key to the analysis is an explanation of the residual identification risks resulting from a synthetic dataset, and how these must be understood in practical terms. The legal analysis addresses the delineation between personal data and anonymous data for synthetic data in the relevant legal sources. The provisions of the GDPR's are intended to safeguard human rights, and the article considers human rights and what weight they may have in the categorization of synthetic data and how these can be processed. An example of the benefit of synthetic data is that they enable the development of diagnostic tools for rare diseases such as childhood cancer, which has implications for children's rights. If the processing of synthetic data for the development of AI models can be done with less restrictions, it will facilitate data access, reduce administrative costs for society and enable the development of AI models for healthcare at a faster pace.

---

[10] Regulation (EU) 2022/868 European data governance (Data Governance Act), recital 7.
[11] Andrew Yale et al., 'Assessing privacy and quality of synthetic health data' (Proceedings of the Conference on Artificial Intelligence for Data Discovery and Reuse, 2019).
[12] PHG, *Are synthetic health data 'personal data', PHG foundation* (Series Title, 2023).
[13] Regulation (EU) 2024/2689 Laying down harmonised rules on artificial intelligence (AI act) , art. 2(7).

The resulting legal assessments of synthetic data are discussed and used to conclude on the empirical case and provide generic reflections on the field. Finally, the last section of the article summarizes the conclusions and discuss them in a legal policy perspective to identify possible needs for clarifications in EU law within sectors that have specific needs. In this chapter, the method is of a legal and political nature, evaluating the necessity to counteract ambiguities and achieve societal benefits.

## 2 Methods and legal sources

### 2.1 Methodological approaches and challenges

This article applies three interrelated methodological approaches to explore the research question: empirical case analysis, legal dogmatic method and legal policy evaluation. The legal analysis provides the foundation for the empirical research process. The conclusions derived from empirical research are then employed in the legal subsumption and in the legal policy discussions.

One of the key challenges in regulating new technologies is their evolving nature, which necessitates continuous assessments of risk, benefit and uncertainty. The empirical study conducted in this article highlights the practical challenges of privacy attacks on synthetic data and the real-world risks of sensitive information disclosure when such data made public. A detailed description of the empirical study is provided in Appendix C. The evaluation framework developed through the legal analysis is applied on the results of the empirical case. While context-specific parameters that determine the risk prevent direct generalization, the assessment method and general principles remain applicable to other synthetic data cases.

The legal dogmatic method involves a systematic examination of legal sources, starting with the definition of personal data in GDPR art. 4. Legal sources are analysed to identify relevant considerations, followed by a subsumption focusing on how the design and function of synthetic data influence the risk of identification of data subjects. Given the complexity of the legal landscape, it is essential to determine which legal provisions are relevant and how they interact. The classification of

Processing of synthetic data in AI development for healthcare and the definition of personal data in EU law synthetic data is not explicitly addressed in existing regulations, making it dependent on legal interpretation.

Fundamental principles in law gain empirical justification when they appear as common denominators across multiple legal systems. Within the healthcare sector, key principles include autonomy, beneficence, prudence ('first, do no harm' – primum non nocere) and technology neutrality.[14] These principles guide the selection of applicable legal rules and help identify potential conflicts.

Human rights form the foundation of EU legislation. For instance, the GDPR preamble (Recital 4) states that 'the processing of personal data should be designed to serve humanity.' EU member states uphold fundamental rights through national constitutions and courts. Proportionality assessments play a central role in application of EU law, in international human rights conventions and in national legal frameworks. These assessments weigh risks against benefits and aim to balance conflicting rules as emphasized in the GDPR rec. 4 (2) stating that data protection is not an absolute right but must be balanced with societal needs and other fundamental rights.

A uniform interpretation of EU law is a prerequisite for the EU single market and requires precise legal wording with consistent interpretations across member states.[15] Conversely, rapid technological advancement that outpaces legal development requires dynamic legal norms. Increasingly extensive uses of new technologies lead to what has been labelled 'the risk society', and its regulation through public law at times leads to unpredictable consequences.[16]

Purpose-driven regulations and flexible legal standards that indicate justification and direction are intended to maintain fairness and rationality in case of unforeseen technological effects post-legislation but leaves greater room for divergent interpretations. EU regulations intended to ensure free movement and competition law have created significant forms of interconnection and interdependencies between public and private law and the economy, and also between individual rights and regulatory regimes.[17]

---

[14] Anne Kjersti Befring, *Health law (original title: Helseretten)* (Cappelen Damm 2022), 2.4.2 p. 79.
[15] European Commission, 'What is the rule of law?' (2024) <https://commission.europa.eu/strategy-and-policy/policies/justice-and-fundamental-rights/upholding-rule-law/rule-law/what-rule-law_en> accessed 18.11.2024.
[16] Gunther Teubner, 'Reflexives Recht: Entwicklungsmodelle des Rechts in vergleichender Perspektive' ARSP: Archiv für Rechts-und Sozialphilosophie/Archives for Philosophy of Law and Social Philosophy 13, 1982; Duncan Kennedy, 'Three globalizations of law and legal thought: 1850–2000' in David M Trubek and Alvaro Santos (eds), *The new law and legal development* (Cambridge University Press 2006); Helmut Willke, 'Ironie des staates: Grundlinien einer staatstheorie polyzentrischer gesellschaft', 1992.
[17] Inger Johanne Sand, 'Globalization and the Transcendence of the Public/Private Divide-What Is Public Law under Conditions of Globalization?' in Claudio Michelon Cormac MacAmlaigh, Neil Walker (ed), *After Public Law* (Oxford University Press 2013) .

Processing of synthetic data in AI development for healthcare and the definition of personal data in EU law

## 2.2 Legal landscape and relevant sources

The AI Act includes several references to synthetic data. These will come into play when synthetic data is used to develop or deploy high-risk AI and general AI, providing requirements for data governance and transparency.[18] In terms of processing personal data, the AI Act art. 2 (7) refers to the GDPR. Synthetic data is not explicitly mentioned in the GDPR, so to determine whether synthetic data qualifies as personal data under GDPR art. 4(1), the legal analysis begins with the wording and recitals and proceeds to investigate the intent and rationale behind the definition.

EU case law has established that in order to interpret provisions of EU regulation, it is necessary to consider not only the wording, but also the context in which it occurs and how the objectives of the rules are expressed.[19] The historical context of the regulation can provide insight into the legislative intent, yet in evolving areas of society it may lose importance over time compared to the broader objectives of the regulation. This legal analysis incorporates multiple sources to assess whether synthetic data meets the legal definition of personal data and identifies key determining factors including GDPR with recitals, EU case law, national case law, guidelines and opinions.

Case law from the Court of Justice of the European Union (CJEU) on synthetic data derived from health data is limited. This study therefore mostly relies on CJEU rulings that clarify the boundary between personal and anonymous data according to GDPR art. 4(1) and that apply a risk-based approach to understanding the GDPR scope. Judgments from the CJEU enriched with opinions on the cases by the Advocate Generals offer valuable guidance on how the regulation should be applied to synthetic data. Since CJEU rulings are binding on Member state national courts, they carry significant legal weight in defining whether synthetic data falls under the GDPR. Several key EU cases are analysed to assess whether synthetic data should be treated as personal data according to the legal definition.[20] In addition, a few national judgements are included that may be relevant to consider.[21]

---

[18] AI Act , recital 111: Documentation of computational footprint for classification of general-purpose AI models, recital 133: transparency and marking of AI generated content, art. 10: bias detection. In art. 50 para 2, transparency requirements for synthetically generated image, video, audio and text data are regulated.
[19] C 487/21 *Österreichische Datenschutzbehörde and CRIF* [2023] ECLI:EU:C:2023:369 CJEU, para 19.
[20] CJEU cases C- 582/14 *Patrick Breyer v Bundesrepublik Deutschland* [2016] ECLI:EU:C:2016:779 CJEU; C-604/22 *IAB Europe v Gegevensbeschermingsautoriteit* [2024] ECLI:EU:C:2024:214 CJEU; C-683/21 *Nacionalinis visuomenės sveikatos centras* [2023] ECLI:EU:C:2023:949 CJEU;C- 319/22 *Gesamtverband Autoteile-Handel e.V. v Scania CV AB* [2024] ECLI:EU:C:2024:103 CJEU; C- 479/22 *OC v European Commission (2024)* [2024] ECLI:EU:C:2024:215 CJEU. Also relevant, the *Judgment of the General Court, T-557/20 - Single Resolution Board v European Data Protection Supervisor (2023)* and appeal case with *2025 Opinion of Advocate General Spielmann regarding case Case C-413/23 P European Data Protection Supervisor v Single Resolution Board (Appeal case of the Banco Popular case)* [2025] ECLI:EU:C:2025:59 CJEU.
[21] VG Hamburg- 21 K 1802/21- ECLI:DE:VGHH:2022:0728.21K1802.21.00 (The Administrative Court of Hamburg 2021); 'Rb. Overijssel- Zwo 22/775- ECLI:Nl:Rbove:2024:594', (Overijssel District Court 2024).

Processing of synthetic data in AI development for healthcare and the definition of personal data in EU law

EU law consists of treaties (primary law) and legal acts (secondary law) and takes precedence over member state national law if a conflict arises (principle of primacy). Regulations must be considered in both *horizontal* and *vertical* legal perspectives.

The *horizontal* is reflected in how EU regulations build upon and incorporate one another and how regulations at the same level may overlap.[22] The AI Act art. 2 (7) explicitly references GDPR for the provisions regarding processing of personal data, so that no conflicts arise between the two. If synthetic data is deemed personal data, it falls under GDPR restrictions. Within the *vertical* perspective the hierarchy of norms cascades downwards. International law and human right treaties take precedence over EU primary law, EU secondary law, national regulations and, finally, soft law like non-binding recommendations, standard and opinions. (see Figure 1).

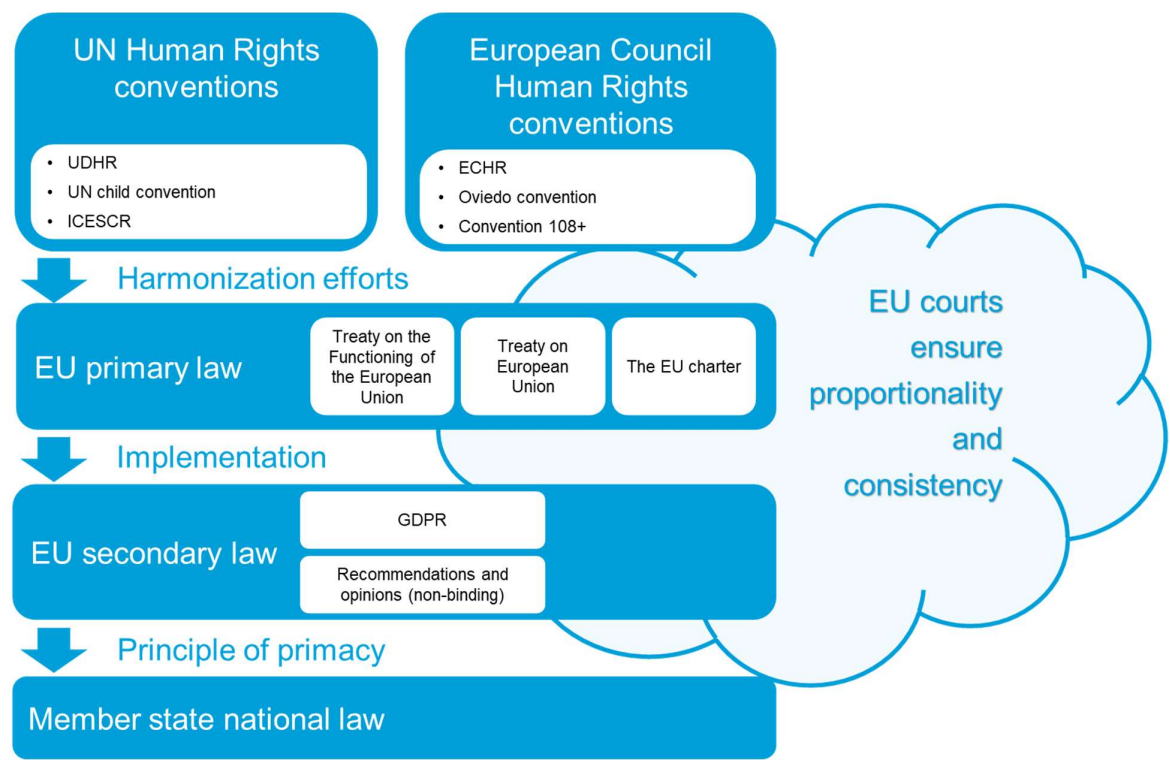

*Figure 1 Relation of primary and secondary law in EU and the relevant sources.*

*Alt text: The image illustrates the layers in the hierarchy of EU law. It has Human rights conventions at the top with an arrow of harmonization efforts pointing down to the EU primary law layer. An arrow of implementation points to the next level below, which is EU secondary law. From there, the last arrow with principle of primacy points to Member state national law.*

---

[22] Anne Kjersti Befring, 'Governance and The EUs AI Act as part of a complex framework. I: Grasping the Complexity of Law. Karnov Group Norway AS 2024 ISBN 9788293816836. p.113-139', 2024.

## Processing of synthetic data in AI development for healthcare and the definition of personal data in EU law

Interpretation of international law follows the Vienna Convention on the Law of Treaties (VCLT), art. 31.[23] The general rules on the sources of law in international law are laid down in the Statute of the International Courts of Justice (the PCIJ Statute, then the ICJ Statute).[24] Art. 38 (a) to (c) of the ICJ Statute defines treaties, customary international law and general principles of law recognized by civilized nations. Many member states are signatories to international human rights conventions and subsequently, EU primary law is harmonized with the Council of Europe and UN human rights conventions making them binding for the EU Member States.[25] There are examples where the harmonization concern between the EU and the Council of Europe goes the other way. When the Council of Europe's Data Protection Convention (Convention 108) was revised and modernized, one of the aims was to reflect the EU's GDPR.[26] This article assumes that human rights are harmonized without further problematization or detailing, although there may be differences between human rights in EU, UN and Council of Europe regulations, as illustrated in the overview in Figure 1.

The EU has its own catalogue of human rights in primary law. The Treaty on the Functioning of the European Union (TFEU) outlines the structure, powers and functioning of the EU and addresses the EU's fundamental principles and rules.[27] The Treaty on European Union (TEU) builds on human rights principles and can be seen as an instrument to operationalize these.[28] TEU art. 2 establishes human dignity, democracy and the rule of law as core values.[29] As stated in the title, the Charter of Fundamental Rights of the European Union (EU Charter) contains the fundamental rights.[30] The TFEU art. 16 and the TEU art. 6 (1) state that the EU Charter has 'the same legal value' as the EU Constitutional Treaties.

The Universal Declaration of Human Rights (UDHR) of 1948 forms the basis for the adoption of human rights conventions by the Council of Europe and the UN.[31] When assessing the legal framework relevant for synthetic data, there are several other international human rights conventions that may also be relevant. The European Convention on Human Rights (ECHR) of 1950 obliges the countries that have acceded to the Convention to respect human rights.[32] The Council of Europe has its own court in the European Court of Human Rights (ECtHR). The UN International Covenant on Economic, Social and Cultural Rights (ICESCR) of 1966 and the UN Convention on the Rights of the Child are of great importance in the field of health.[33] The Biomedical Convention (Oviedo Convention) of 1996 is the only international instrument that specifically regulates human rights in the biomedical field.[34] Although international courts and the EU Court are distinct judicial systems, the principle of proportionality will remain a consistent factor. While the specific judgements in the EU court cases that are relevant to the analysis in question do not explicitly address the international conventions to balance interests, they apply the principle of proportionality.

---

[23] *Vienna Convention on the Law of Treaties (VCLT).*
[24] League of Nations, *Statute of the Permanent Court of International Justice* (16. december 1920), article 36 (a); United Nations, *Statute of the International Court of Justice* (26. juni 1945), art. 38; VCLT articles 31–33.
[25] This is expressed in C- 540/03 *European Parliament v Council of the European Union* [2006] ECLI:EU:C:2006:429 CJEU, pargraph 37, which states that the Court may 'point out that the International Covenant on Civil and Political Rights is one of the international instruments for the protection of human rights which it takes into account in applying the general principles of Community law.' In the Opinion of Advocate General Sharpston in C- 73/08 *Bressol and Others* [2010] ECLI:EU:C:2010:181, CJEU, para 126, reference was made to the relevance of the International Covenant on Civil and Political Rights (ICCPR) in the application of the general principles of Community law in EU law.
[26] Convention for the protection of individuals with regard to the processing of personal data (Convention 108 +).
[27] Consolidated version of the Treaty on the Functioning of the European Union (TFEU).
[28] Consolidated version of the Treaty on European Union (TEU).
[29] TEU, art. 2.
[30] *The Charter of Fundamental Rights of the European Union (EU Charter)* (OJ C 202/02 edn, 2016).
[31] The Universal Declaration of Human Rights (UDHR).

EU regulations must be general enough in nature to be robust for societal, legal and technological developments. Specific standards and soft law instruments contribute an important role clarifying interpretations within different specific fields. However, soft law is not legally binding but in the event that the courts attach significant weight to them, they can assume a role as a legal source.

The article analyses soft law (non-binding instruments) and legal theory to assess whether there are any significant legal arguments for how synthetic data should be categorized. In this context, opinions from the European Data Protection Supervisor (EDPS) and the European Data Protection Board (EDPB) are relevant. Guidance from the EDPB is intended to ensure a consistent application of the GDPR, replacing the Working Party (WP29) established under art. 29 of the 95/46/EC Data Protection Directive (DPD).[35] Some opinions from the WP29 are discussed in the article as they can still be considered relevant for taking a position on the categorization of synthetic data.

## 3 Empirical case study to test identifiability of a synthetic dataset
### 3.1 The synthetic dataset in the empirical case

In this case study a practical test – a Membership Inference Attack simulation- was applied to investigate whether synthetic data could be used to infer membership of patients in the original dataset that was used for training the synthetic data generator. For detailed descriptions of the experiment, see Appendix C. Synthetic data generation involves creating artificial data that mimics the statistical properties of real data. It is typically generated programmatically using algorithms or computer simulations and is defined as artificially generated rather than captured directly from real-world events.[36]

---

[32] The Council of Europe, European Convention of Human Rights, as amended by protocols Nos. 11,14 and 15 (ECHR) of 2021.
[33] United Nations, *International Covenant on Economic, Social and Cultural Rights (ICESCR)* (1966)  The UN Convention on the Rights of the Child (1989).
[34] The Council of Europe, Convention on Human Rights and Biomedicine (The Oviedo Convention) of 1997.
[35] EDPB, 'Endorsed WP29 Guidelines'   <https://www.edpb.europa.eu/our-work-tools/general-guidance/endorsed-wp29-guidelines_en> accessed 18.11.2024.

In this empirical case, a synthetic dataset was generated based on cancer patient data from the Florence project—an interregional collaboration utilizing cancer registry data to improve clinical decision-making in colorectal cancer.[37] The dataset comprises 16 attributes describing the diagnosis and treatment of 30.631 colorectal cancer patients, sourced from the Norwegian Cancer Registry. Direct identifiers were removed from the dataset and other variables were modified to reduce granularity, as a combination of detailed information can potentially be used to identify a person when combined with additional information. El Emam, Mosquera and Bass call these 'quasi-identifiers'.[38]

Different types of synthetic data pose varying privacy risks depending on how they are generated. Synthetic data can be derived from explicit knowledge, such as the laws of physics in simulation models, which inherently carry no risk of personal data exposure. Conversely, synthetic data generated from real-world datasets may be vulnerable to inference attacks.[39] Within this category, one can distinguish between:

- *One-to-one derived* synthetic data, which is directly derived from real data through a transformation process and maintains a one-to-one relationship with original data points.
- *Real data inspired* synthetic data, which replicates statistical distributions without preserving exact individual records.

Figure 2 illustrates the different types. This distinction is crucial in assessing privacy risks and the potential for re-identification in different synthetic data applications.

---

[36] Alloza et al., 'A Case for Synthetic Data in Regulatory Decision-Making in Europe', 2023 (n 7).
[37] Florence, 'The Florence Project home page' <https://florence.forskning.eu/en/> accessed 19.11.2024.
[38] Khaled El Emam, Lucy Mosquera and Jason Bass, 'Evaluating identity disclosure risk in fully synthetic health data: model development and validation' 22 Journal of medical Internet research e23139, 2020.
[39] Samuel Yeom et al., 'Privacy risk in machine learning: Analyzing the connection to overfitting' (2018 IEEE 31st computer security foundations symposium (CSF), 2018)

Processing of synthetic data in AI development for healthcare and the definition of personal data in EU law

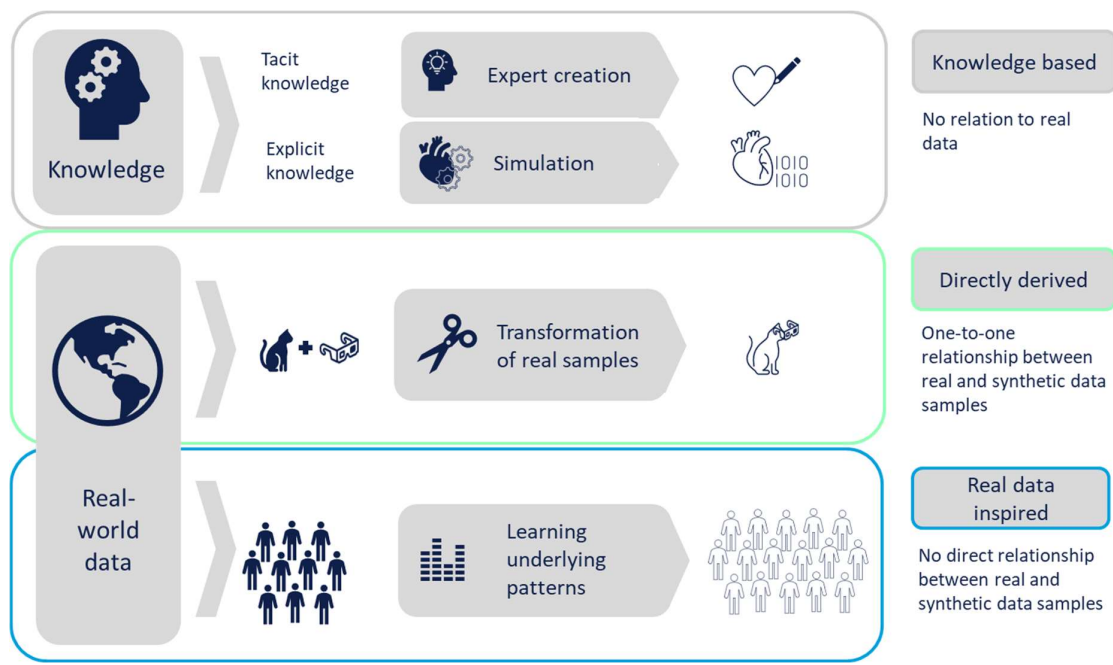

*Figure 2 Types of synthetic data, illustration adapted from Vallevik, Marshall, Babic and Nygård.[40]*

*Alt text: The image shows the spectrum of different types of synthetic data and is adapted based on Vallevik, Marshall, Babic and Nygård 2025. The image shows how synthetic data can be made either from knowledge or real data and can be synthesized through different techniques. The types based on real data can be either directly derived by transforming the data, or it can be inspired – by learning the underlying distribution.*

Directly derived synthetic data may be at risk of reverse engineering to identify persons in the original dataset. While the 'real data inspired' synthetic data do not have a one-to-one relationship to original data, the generators are commonly built to optimize similarity to the training dataset, at the risk of creating data points that are materially similar to the original. This means that even though deep generative synthetic data does not allow a one-to-one relationship between the real personal data in the training data set and the synthetic data points, there may still be a risk of inferring information about the training data from the synthetic data.[41]

This study focuses on real data-inspired synthetic data that is produced using a machine learning-based generator. These are often referred to as deep generative methods or generative AI when deep learning is applied. The synthetic data generator (a Variational Autoencoder) was trained on the Florence dataset to learn its underlying statistical distributions and relationships. Once trained, the model generates synthetic datasets by sampling from these learned patterns. This results in data that preserves statistical characteristics of the original dataset without directly replicating individual records as illustrated in Figure 3.

---

[40] Vibeke Binz Vallevik, Serena Elizabeth Marshall, Aleksandar Babic, Jan Franz Nygaard. , 'Opinion: Rethinking Synthetic Data definitions- A privacy driven approach ' Nordic Machine Intelligence 2025.
[41] Chang Sun, Johan van Soest and Michel Dumontier, 'Generating synthetic personal health data using conditional generative adversarial networks combining with differential privacy' Journal of Biomedical Informatics 104404, 2023.

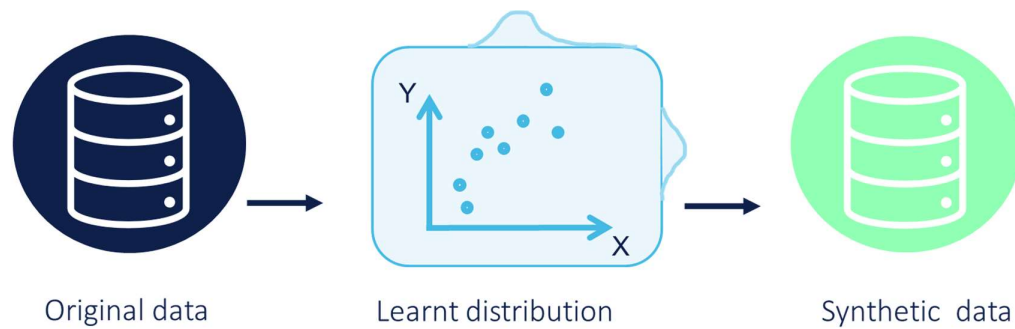

*Figure 3 Generation of synthetic data based on learning the statistical distributions of an original dataset.*

*Alt Text: A simplified diagram shows the idea of how data flows from an original training dataset through a learnt distribution inside the synthetic data generator and on to a synthetic dataset. The illustration emphasizes how there is no one-to-one or direct relationship between the training data and the synthetic data.*

### 3.1 Measuring the identification risk through a Membership Inference Attack

When a synthetic data generator creates new data based on distributions learned from training on real-world dataset, there is a risk that it may accidentally produce a close or even exact match to an original data record. If the synthetic data closely resembles the original data statistically, it could hold a residual risk of identification. There are several approaches to measuring identifiability risk in synthetic datasets.

The number of exact matches can be measured and will depend on the generation method and the nature of the dataset. A high number of exact matches in a heterogeneous dataset could indicate that this statistical closeness may leave the dataset vulnerable to inferences of the identity of persons in the original dataset.

Measuring statistical closeness between the original dataset and the synthetic dataset can give an indication of how robust the dataset is for inferencing. Most commonly used is an aggregate estimate measuring the distance of 'Nearest Neighbours' between datapoints in the synthetic and original dataset.[42] These statistical metrics demand low computational power but do not give an absolute answer to the likelihood of a person being identified.

---

[42] Matteo Giomi et al., 'A unified framework for quantifying privacy risk in synthetic data' (23rd Privacy Enhancing Technologies Symposium, 2023).

Processing of synthetic data in AI development for healthcare and the definition of personal data in EU law

Other methods include differential privacy which uses mathematical 'guarantees' (ε) in the form of added errors in the generation methods.[43] This defines the intended privacy budget well if the intent is to compare generation methods, but does not provide clues as to the actual residual identifiability in the synthetic data.[44] The risk of confidentiality breach is dependent not only on the model, but also on the nature of the dataset itself – size, number of variables, diversity and rare samples. Many similar individuals in the dataset and few differentiating variables makes identification more difficult or impossible. Access to additional information regarding generation method, model or hyperparameters may make the attack more precise. If a similar open dataset exists that can be used to check the inference model's performance, this can be an indicator for the attacker regarding the success of their attack model.

Substantially higher computational power is needed to investigate the actual risk in more practical terms, through simulation of different forms of inference attacks. The most common ones are Membership Inference Attacks (MIA) and Attribute Inference Attacks (AIA). The MIA was introduced by Shokri et al. in 2017 and is a shadow modelling technique to simulate an attacker that attempts to identify whether specific individuals were 'members' in the dataset that was used to train the synthetic data generator, see Figure 4.[45] Inferring membership can be a step towards other privacy breaches like profiling or attribute inferences.[46] AIA is built on the same logic as the MIA, using shadow modelling to infer sensitive or private attributes about specific individuals. The empirical case uses MIA to investigate the actual practical risk of identifying individuals from the training data that was used to train the synthetic data generator.

---

[43] James Jordon, Jinsung Yoon and Mihaela Van Der Schaar, *PATE-GAN: Generating synthetic data with differential privacy guarantees* (2018) .
[44] Andrea Gadotti et al., 'Anonymization: The imperfect science of using data while preserving privacy' 10 Science Advances eadn7053, 2024 .
[45] Reza Shokri et al., *Membership Inference Attacks against machine learning models* (IEEE, 2017) .
[46] Emiliano De Cristofaro, 'A critical overview of privacy in machine learning' 19 IEEE Security & Privacy 19, 2021

Processing of synthetic data in AI development for healthcare and the definition of personal data in EU law

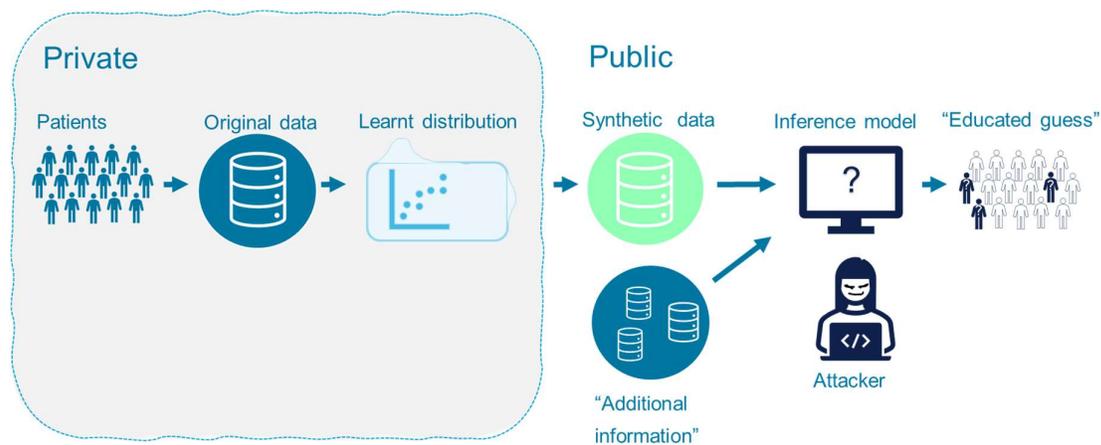

*Figure 4 The structure of a Membership Inference Attack. It is distinguished between available information from the data processor and public information available to a third party.*

*Alt text: The image illustrates the data flow in a Membership Inference Attack. The Private sphere is depicted with a box on the left, which is where the original dataset is. The synthetic dataset is illustrated to be published on the outside. This, and another box of "additional information" feeds into an inference model, where an attacker is depicted making educated guesses about the patients in the original dataset.*

In the context of synthetic data, the goal of an MIA is to determine if the data from a specific individual (i.e. the target record) was used to train the generator model that created the synthetic datasets. If the attacker identifies that the target record contributed to this model, sensitive information may be exposed. For example, discovering that a particular individual's health record was included in a dataset of cancer patients could reveal protected information that should remain confidential. The attacker could have a specific individual – a natural person – that they want to try to identify in the dataset or could take a shotgun approach and try to map the dataset with a wider population to see whether they could identify someone.

The vulnerability in a released synthetic dataset towards an MIA can be simulated as a privacy game between a challenger and an attacker. To instantiate the game, the challenger releases a synthetic dataset where a certain target record was either included or excluded during the data generation process. The attacker is then challenged to infer whether the target record was used in generating the synthetic dataset by leveraging the available information.

## 3.2 Attack setup and performance

The assumed attack scenario should be built on the type of information available to an attacker. The attacker will have access to a synthetic dataset and will likely be able to access a certain amount of additional information to aid in the identification. This can be knowledge of or access to the generator, method and hyperparameter setup, access to a dataset similar to the original dataset or knowledge of parts of the real data, but not the original dataset itself. The MIA setup is based on what is considered a more realistic scenario of what available information the attacker is likely to be able to access. The additional information in the hands of the attacker could be anything that is considered common knowledge, like age or gender, but also information gathered through social media platforms. For example, it could be a social media status update disclosing information about a medical incident like a diagnosis or an intervention like for example surgery, which also provides an approximate point in time for the incident. Medical codes or detailed clinical information from the medical records is not considered open or common knowledge and therefore not considered available information in this scenario.

A recent review of anonymization and privacy techniques highlights publication bias in studies of Membership Inference Attacks against synthetic data.[47] Many of these experiments adopt overly conservative assumptions where the attacker is granted unrealistic levels of knowledge—such as for example perfect access to the target record from the original training dataset. Unsurprisingly, such setups tend to overstate the identification risk. This does not reflect a practical or realistic scenario and will lead to overestimation of the attack effectiveness. To check for sensitivity, this experiment included this kind of control scenario where all variables were available – as if the attacker had access to the original training data- – to shed light on the absolute outer bounds of the risk and not to be viewed as a likely scenario.

The attack models require several design choices affecting their efficacy and efficiency. Fine-tuning these parameters requires access to the original data, which would not be available for an attacker. Through the simulation study, different success rates are illustrated and indicate practical challenges an attacker will face when implementing an attack.

The simulation of an MIA was performed on the Colossus high-performance computing cluster hosted by the University of Oslo, reserving 300 GB memory and 20 cores. The first setup took more than the maximum allowed week to run on this infrastructure. The experiment had to be re-designed to fit the infrastructure limitations. Through the Colossus HPC system, 300 GB of memory and 64 cores were requested to parallelize the rounds of the attack simulation. Each run of the simulation took up to two hours to complete. The whole process took approximately 6 weeks to set up and perform.

### 3.3 Resulting identification risk

The result of a simulation attack is the performance metrics of the inference model (see Figure 4) which is not directly transferable to a 'probability of identification'. The success of an attack simulation is measured by comparing the attacker's guess with the ground truth – the actual inclusion or exclusion of a target record. By repeating the privacy game multiple times, alternating between including and excluding different target records, the simulation is aimed at assessing the overall vulnerability of the synthetic dataset. Since not all target records are equally vulnerable, the likelihood of a successful attack can vary depending on which record is selected as the target record.[48]

---

[47] Gadotti et al., 'Anonymization: The imperfect science of using data while preserving privacy', 2024 (n 44)44.

The performance of an attack simulation can be summarized though a confusion matrix that consists of four metrics: True Positives (TP), True negatives (TN), False Positives (FP) and False negatives (FN), see Figure 5.

*Figure 5 Confusion matrix illustrating performance of an attack simulation. The attacker's guess is the resulting guess from the inference model used for the attack.*

*Alt text: A confusion matrix is illustrated as four squares summarizing different performance metrics from a classification model. The four types of outputs can be either True Positive, False Negative, False Positive or True Negative.*

A True Positive result means the attacker correctly guessed the record was included in the training dataset. True Negative means the attacker correctly guessed the record was not included. False Positive means the attacker guessed the record was included, when it was not. False Negative means the attacker guessed it was not included, when it was. From these four metrics, most performance measures can be calculated, among others the True Positive Rate (TPR) and False Positive Rate (FPR). The True Positive Rate is the number of true positives divided by the total number of included records, the top row in the confusion matrix of Figure 5:

$$TPR = \frac{TP}{(TP + FN)}$$

*False Positive Rate* is the FP over the total number of records that were not included, the bottom row of the confusion matrix:

$$FPR = \frac{FP}{(FP + TN)}$$

---

[48] Matthieu Meeus et al., 'Achilles' heels: vulnerable record identification in synthetic data publishing' (European Symposium on Research in Computer Security, 2023).

Typically, a performance score is used to summarize the outcome of an attack simulation, comparing the known outcome to the attacker's guess. Quite common is the use of the Receiver-operating characteristic curve (ROC) Area Under Curve (AUC) score, see Figure 6. The ROC curve is made by graphing the true positive rate (TPR) over the false positive rate (FPR), and the AUC is the area under this curve and is given as a number between 0 and 1. An AUC score close to one indicates a perfect attack, while 0.5 reflects random guessing. Lower than 0.5 typically signifies that the inference method employed by the attacker to indicate target record membership is inappropriate due to a suboptimal attack configuration.

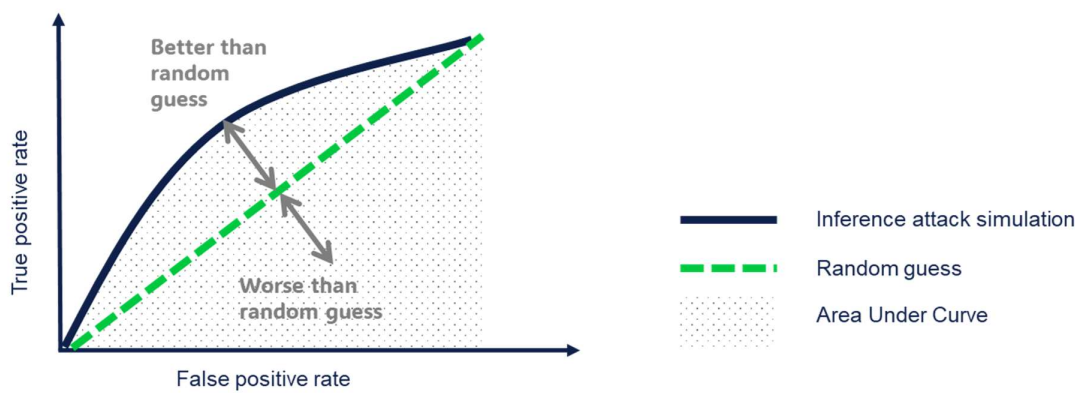

*Figure 6 ROC for a simulation versus random guesses. The AUC is the shaded area under the curves. AUC of 1 would mean all the guesses are correct, while an AUC of 0.5 means the guesses are completely random.*

*Alt text: A graph illustrating the ROC with its Area Under Curve (AUC). The x-axis on the graph is False positive rate, the y-axis is true positive rate. The line of a random guess – along the 0,5 line is depicted. The graph shows that any AUC score above 0.5 is better than a random guess, while below 0,5 is worse that a random guess.*

When academic researchers report that they found privacy risk in synthetic data, it usually means they get a ROC AUC score higher than 0.5 meaning that the inference model provides a better guess than a random guess. What this means in practical terms of actual identifiability risk is most commonly not discussed.[49]

The attack simulation in the empirical case is run with two different scenarios: an *Upper Bounds* scenario and a *Likely Scenario*. The basis for the analysis is the *Likely scenario*, while the scenario called *Upper bounds* is included only to illustrate the delineation to the upper bounds of possible results, being a scenario in which the attacker has access to all information about a patient (target record).

---

[49] Gadotti et al., 'Anonymization: The imperfect science of using data while preserving privacy', 2024 (n 44)44.

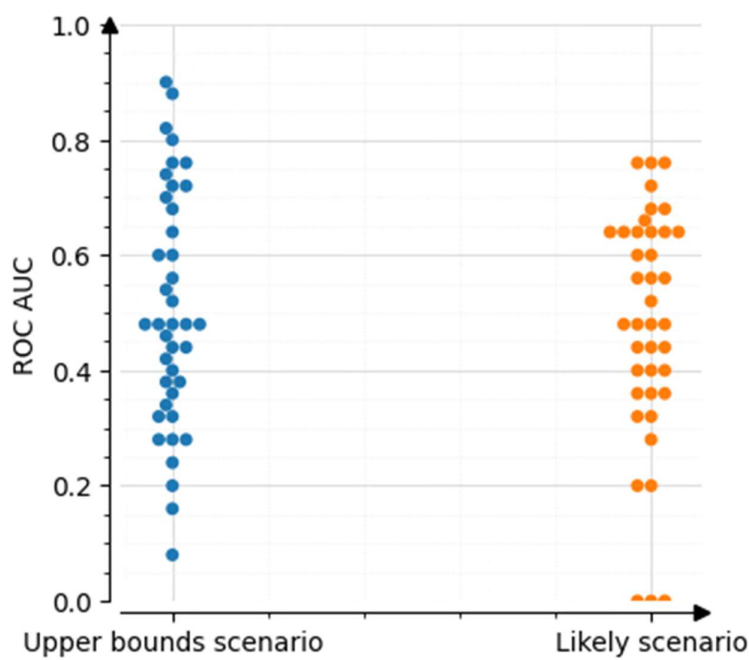

*Figure 7 Swarm plot of the attack simulation results showing the ROC AUC scores for different target records in three different attack scenarios.* Likely scenario *is the basis for the analysis, while the other illustrates the upper bounds of the risk.*

*Alt text: The graph shows swarm plots of AUC results from the empirical case. On the left is the Upper bounds scenario, which is spread all the way from 0.1 to 0.9. The likely scenario, to the right shows results all the way to 0 and max up to 0.75.*

Figure 7 shows each dot in the plot corresponds to multiple guesses for a single patient and the ROC AUC score for a specific attack. In the likely scenario, there are no records with a score higher than 0.75.

However, the use of performance scores to measure an attacker's success and the vulnerability of the data generation process is only feasible when both the ground truth and the attacker's guess are available. Outside of a simulation setting, only the holder of the original data has access to the ground truth and can verify how well the inference model is performing. Although the attacker can have strong indications about the result, it is not possible to confirm the suspicion without actually consulting the training dataset itself.

## 4 Synthetic data as personal data regulated by the GDPR or as data outside the GDPR

The relevant elements for the legal dogmatic analysis are presented in this section, prior to the analysis where the practical identifiability will be discussed in light of the legal conclusions.

Processing of synthetic data in AI development for healthcare and the definition of personal data in EU law

## 4.1 The definition of personal data in GDPR

The processing of synthetic data classified as personal data will fall within the material scope of GDPR Art. 2 (1), as specified. This also applies to the processing of personal data which is wholly or partly automated, or which forms part of a 'filing system or is intended to form part of a filing system other than by automated means.'

There must be a basis for lawful processing of personal data according to GDPR art. 6. In healthcare, data concerning health is a special category of personal data in GDPR art. 9(1) and can only be processed lawfully according to the exceptions in GDPR Art. 9 (2).

In GDPR recital 26 (5-6), there is a clarification that the GDPR and the principles of data protection do not apply to anonymous information, including personal data that has been anonymized in such a way that the data subject is no longer identifiable. This statement assumes that data can be anonymized and processed without being covered by the GDPR, including 'for statistical or research purposes.'

The definition boundary between personal data and anonymous data is therefore of great importance for whether and how synthetic data can be processed.

When assessing whether synthetic data is personal data, a natural starting point is the definition of personal data in GDPR Art. 4 (1), which assumes that data can be linked to a natural person either directly or indirectly, and as it states:

> '"personal data" means any information relating to an identified or identifiable natural person ("data subject"); an identifiable natural person is one who can be identified, directly or indirectly, in particular by reference to an identifier such as a name, an identification number, location data, an online identifier or to one or more factors specific to the physical, physiological, genetic, mental, economic, cultural or social identity of that natural person;'

The provision specifies that it applies to a natural person which is an individual human being, a physical person. This is different from a digital avatar or a legal person, which can be a company. GDPR does not apply to the personal data of deceased persons but has a provision for member states to provide rules for deceased persons.[50]

Direct identification means that the information can be linked to a natural person through identifying factors mentioned in art. 4 (1) such as name, identification number etc. The concept of 'any information' in the definition provides a wide scope, only restricted by the fact the information must relate to a natural person and identify them or make them identifiable. Consent strings, tax information and specific biometric data such as images, audio and video are personal data if they can be linked to a natural person and thus enable identification.[51]

---

[50] GDPR, recital 27 (1-2).

The concept of identifiability is elaborated in recital 26 (3), where the boundary between personal data and data that are anonymous is linked to 'all the means reasonably likely to be used, such as singling out, either by the controller or by another person to identify the natural person directly or indirectly'. Means are referred to as considerations that must be assessed. This phrasing emphasizes that the identifiability test is essentially risk-based and requires assessment of not just the possible means but whether they are likely to be used.

Recital 26 (4) establishes that for assessing what is 'reasonably likely', all objective factors should be considered. Such factors are the resources required to carry out the identification of persons in a synthetic data set, like costs involved and the time it takes.

The likelihood of a successful identification attempt from synthetic data can be expected to influence the motivation to invest resources into identifying someone from the dataset. Therefore, understanding the capabilities of current and expected future technologies and the associated efforts needed for re-identification is crucial. The required competence for an identification attack could also be considered an investment or cost factor.

Synthetic data is not explicitly mentioned in the GDPR definition or preamble, and only one CJEU case- the 'Covid tracking case'- specifically addresses whether 'fictitious data' can be personal data under the GDPR[52]. The case concerned the processing of IT test data that was based on copies of real data, but fictitious data was created for testing purposes. The Court states that fictitious data that does not relate to an identified or identifiable natural person, but rather to a person who does not actually exist, does not constitute personal data within the meaning of art. 4(1) of the GDPR.[53] Conversely, fictitious data that relate to a natural person is personal data. The Opinion from Advocate General Emiliou in the same case provides increased detail, as he emphasises that information can be regarded as personal data regardless of whether it is true or false information, as long as it is related to a natural person.[54]

Synthetic data that has been created based on existing knowledge rather than based on real-world data would easily fall into the category of fictitious data that does not relate to a natural person. The relevance to this judgement is more unclear for synthetic data that is derived from real data. The patients in the synthetic dataset are fictious yet the dataset may still hold a residual risk of identifying individuals from the original dataset through inference.

---

[51] Christopher Kuner et al., *The EU general data protection regulation: a commentary* (Oxford University Press. 2020).
[52] *C-683/21 Covid tracker case* (n 20).
[53] *C-683/21 Covid tracker case*, para 55.
[54] Opinion of Advocate-Geneal Emiliou on 4 May 2023 in Case *C-683/21 Covid tracker case*, para 28.

## 4.2 Pseudonymous data and the use of additional information to identify data subjects

Closely linked with the concept of identifiability in recital 26 is pseudonymous data, when additional information combined with the dataset make identification possible. Pseudonymization is a technique to reduce identifiability and is defined in GDPR art. 4 (5) as *'…the processing of personal data in such a manner that the personal data can no longer be attributed to a specific data subject without the use of additional information'*. The provision requires that such additional information is stored separately and is subject to technical and organizational measures to prevent the information from being traced back to a specific person.

The traditional approach to pseudonymization is to swap direct identifiers with serial numbers and for the data controller to keep the reidentification key separate.[55] In medical research, this is done to protect the identity of participants for the data processing while allowing for identification of a patient if needed, for example when patients in a clinical study have opted to be alerted about findings of clinical relevance to them or other incidental findings. Other efforts to reduce identifiability are often applied in parallel, such as reducing granularity by generalising certain variables. Synthetic data generated by machine learning do not have reidentification keys, as there is no one-to-one relationship with the original data. It can, however, be discussed whether it should be considered pseudonymous when additional information can be used to infer identification. Recital 26 (2) of the GDPR states that if pseudonymized personal data could be attributed to a natural person, it is still personal data. This suggests that- pseudonymous or not- the issue remains to decide whether the data could be attributed to a natural person.

Recital 26 (3) refers to the agent that has these means to identify the pseudonymous data, saying it can be either '<u>the controller</u>' or <u>'another person'.</u> The controller in GDPR terms is defined by art. 4(7) as *'…the natural or legal person, public authority, agency or other body which, alone or jointly with others, determines the purposes and means of the processing of personal data; (…)'*.

In applying the 'reasonably likely'-criteria of recital 26, several court cases have examined how access to additional information may enable the identification of individuals. These cases will be explored to clarify the definition of personal data in the context of synthetic data generation.

---

[55] ENISA, *Deploying pseudonymisation techniques - the case of the Health Sector* .

## 4.3 Case law that highlights the delineation of anonymity

### 4.3.1 Breyer case

The Breyer case between Patrick Breyer v. Bundesrepublik Deutschland is considered a significant decision by the European Court of Justice regarding the definition of personal data.[56] The case concerns a legal dispute between the German citizen Patrick Breyer and the German government concerning the collection and storage of dynamic IP addresses on the websites of German federal institutions. The provider of online media services- namely the Federal Republic of Germany- recorded IP addresses with the date and time of access to websites. Breyer claimed that the storage of these IP addresses infringed his privacy under German data protection law and argued that dynamic IP addresses should be considered personal data because they could potentially identify him.

The CJEU considered whether the dynamic IP addresses should be considered personal data on the basis of art. 2(a) of the Directive and whether the IP address could 'be attributed to a natural person using additional information'. This case from the CJEU concerned the EU Data Protection Directive 95/46/EC (DPD) and not the GDPR. In connection with the adoption of the GDPR, it was assumed that previous practice would still be relevant.

The Internet provider had additional data which, combined with the IP addresses, would make it possible to identify a person. The Court established that it is not required that all the information enabling identification of the data subject was in the hands of one person for the information to be classified as personal data.[57, 58] This does not, however, mean that the mere existence of additional information that could aid identification automatically makes the IP addresses personal data. The Court considered whether the possibility to combine the data constituted a 'means reasonably likely to be used' by the German Federal Government to link the data subject to IP addresses.[59]

The judgment states that this would not be personal data if identification of the data subject was 'prohibited by law or practically impossible' because it would require a disproportionate effort in terms of time, cost and effort.[60] The assessment of whether the receiver had reasonable means to identify data subjects is based on considering opportunity and effort needed. It introduces the term 'practically impossible' as opposed to impossible, and the notion that legal prohibition is considered to lower the risk of identification. The Court ruled that it would not be reasonably likely that the additional information would be accessed through unlawful activities, as this was not proportional to the gain.

---

[56] *C- 582/14 Breyer*.
[57] *C- 582/14 Breyer*, paras 43-4.
[58] According to the definition in the 95/46/EC Data Protection Directive (DPD) art. 2(a) .
[59] *C- 582/14 Breyer*, para 45.
[60] *C- 582/14 Breyer*, para 46.

The Court established that the online media service provider did in fact have legal means to request the additional information in the event of a cyberattack and thereby also a practical possibility to combine the information that could be used to identify individuals.[61] The IP addresses in question were therefore considered personal data in this case.

The Breyer case concerned the Data Protection Directive that was replaced by the GDPR. In GDPR, recital 26 now specifies these criteria initially set by the CJEU in the Breyer judgement. Reasonable means constitutes both technical means and the effort required for identification in terms of costs and time, taking into account available technology and technological development. The ruling implies that more than a theoretical possibility is required to constitute reasonable means.

### 4.3.2 The Banco Popular case

Another landmark case which is currently under appeal to CJEU in C-413/23 P is the General Court case on the dispute between the Single Resolution Board (SRB) and EDPS, the so-called Banco Popular case.[62] The judgment concludes that pseudonymized data transferred to a data recipient is not personal data if the recipient lacks the means – that is, access to the 'additional information' – necessary to re-identify the data subjects.

The Court relied on the Breyer judgment when it came to the distinction between pseudonymous and anonymous data. This decision clarifies that pseudonymized data can be treated as non-personal data under certain conditions. The dispute arose because the company SRB disagreed with a decision of the EDPS. The SRB had sent a pseudonymized dataset to Deloitte, which contained comments from complaints processed by the SRB. The EDPS considered that the SRB had failed to fulfil its obligations regarding the processing of personal data in the procedure for compensating shareholders and creditors following a resolution of a bank.

The Court found that since Deloitte did not have access to the additional information that could be used for identification, the dataset – in the hands of Deloitte – does not constitute personal data and that this must be assessed in the light of the information available to Deloitte. It is therefore assumed that pseudonymized data can be considered anonymous if the recipient cannot re-identify individuals when considering all means that are reasonably probable the recipient can and will access.

---

[61] *C- 582/14 Breyer*, para 48.
[62] *Judgment of the General Court, T-557/20 - Single Resolution Board v European Data Protection Supervisor (2023)*.

The decision was contested by the EDPB/EDPS and has been appealed to the CJEU.[63] Advocate General Spielmann expresses partial agreement with the GC judgment in his 2025 Opinion.[64] He considers that the EDPS's claim that pseudonymized data is automatically personal data goes too far when they claim a dataset once defined as personal data cannot be anonymous in the hands of another party when the controller is in possession of an identification key. The Advocate General points out that pseudonymization allows for the data subject not to be identified and that the regulation, similarly to the wording in GDPR recital 26, does not apply to personal data that is rendered anonymous in such a manner that the data subject is not or is no longer identifiable.[65] If it is impossible to identify the data subjects, they are therefore considered legally to be sufficiently protected by the pseudonymization process, despite the fact that the additional information that makes identification possible still exists in the hands of the controller. His conclusion is in line with Banco Popular, in that pseudonymous data may or may not be anonymous, depending on whether the pseudonymization '…was sufficiently robust to conclude that the complainants […] were not reasonably identifiable.'[66]

This means that it is not the sum of all existing data as such that is assessed, but the risk that arises when the dataset is in the hands of the recipient and based on the additional information the recipient can reasonably likely access.

### 4.3.3 OC v Commission

The *Case OC v Commission* is often referred to as a landmark case concerning the assessment of the boundary between personal and anonymous data.[67] Although it does not concern synthetic data, it can be used to discuss its classification. The case concerns the identification of the claimant ('OC') by a journalist and the Court uses the GDPR recital 26 test of 'reasonably likely' with the notion of a disproportionate effort. The identification occurred on the basis of a press release which did not contain the name of the researcher, the subject of the research or the name of the project. The press release did describe her gender, nationality, her youth and her profession as a female scientist, together with her fraudulent activities. It contained a reference to her father and the place where he works, as well as the approximate amount of the grant awarded to the researcher, the granting body, the nature of the entity hosting the project and its geographical location. The European Commission held that an average reader could not identify OC based on the information.

---

[63] *2025 Opinion of Advocate General Spielmann regarding case Case C-413/23 P European Data Protection Supervisor v Single Resolution Board (Appeal case of the Banco Popular case)*.
[64] *2025 Opinion of Advocate General Spielmann regarding case Case C-413/23 P European Data Protection Supervisor v Single Resolution Board (Appeal case of the Banco Popular case)*.
[65] Opinion of Advocate General Spielmann, C-413/23 P, para 50-1.
[66] Opinion of Advocate General Spielmann, C-413/23 P, para 59.
[67] *C- 479/22 OC v Commission*.

Processing of synthetic data in AI development for healthcare and the definition of personal data in EU law

The Court examined whether the press release constitutes personal data. The argument of basing the identifiability test on an 'average reader' without close knowledge of the field, was opposed in the judgement. Interestingly, it was also concluded that the fact that a journalist had actually identified the claimant was not sufficient as an argument on its own, the likelihood of identification still had to be considered.[68] The Court considered whether there was a reasonable likelihood that the claimant would be identified based on the press release and any additional information. It held that obtaining the additional information necessary for identification could be done through an internet search, as the university's website provided a list of and information about their funded projects, including the names of project leaders. The ease of identification was assessed against the expected motivation to do so.[69] It assumed that information about fraud or corruption would heighten motivation among both readers and journalists. Given this context, an Internet search by journalists or other persons unfamiliar with her professional background to enable identification means the risk could not be considered insignificant.

The court concluded that identification was practically possible through additional information found online. The effort required was proportionate to the motivation, making it reasonably likely that the person in the press release could be identified. Therefore, the Court determined that the press release constituted personal data.

### 4.3.4 Other relevant cases

The CJEU has several examples of a probability assessment being used when assessing data. In a case referred to as IAB Europe or simply IAB, a company had developed a system to safeguard confidentiality and requirements for necessary transparency 'Transparency & Consent String' (TC string) based on the consent of the data subject.[70] The TC string was shared with advertisers and data brokers. The court pointed out that upon request, IAB Europe members were required to provide IAB Europe with information that would enable them to identify the users whose data was the subject of a TC string.[71] The string, combined with a cookie placed on the user's device by the Consent Management Platform, can reveal the user's IP address. In this scenario, the cookie was 'additional information' that enabled identification. The Court concluded that IAB Europe did not possess the additional information, but that it had reasonable means to acquire it through the right to request it from their members. This was considered sufficient to conclude that art. 4(1) of the GDPR must be interpreted as meaning that a string such as the TC string which was composed of a combination of letters and characters and that contains a user's consent preferences concerning processing of personal data, is in itself personal data if the data could be associated with an identifier, such as an IP address by reasonable means.[72]

---

[68] *C- 479/22 OC v Commission*, para 58.
[69] *C- 479/22 OC v Commission*, para 63.
[70] *C-604/22 IAB*.
[71] *C-604/22 IAB*, para 48.

In the Scania case, the European Court of Justice ruled that vehicle identification numbers can be personal data.[73] Scania, a European manufacturer of heavy vehicles, provided access to repair and maintenance information via a website for mechanics. This website allowed searches for general vehicle details or the last seven digits of the vehicle identification number (VIN), which is a serial number. The trade association of independent wholesalers of vehicle parts, Gesamtverband Auto-handel claimed that this was restrictive of competition and required Scania to provide this access also for independent operators. The Court ruled that the VIN is not personal data in nature but may become personal data under GDPR art. 4(1) when a third party who has access to it has the means to identify the owner of the vehicle.[74] The Court emphasizes that it is the perspective of the data holder and what kind of additional information it is reasonable to assume that they may have access to combine with the information in question, which is decisive for whether it is personal data.

National courts are deciding on the scope of the GDPR, which may contribute to how it is enforced, even if EU law may reach a different conclusion. A relevant judgment from a local German court concerns whether health data in the cancer registry should be considered anonymous.[75] The case arose when a data subject requested access to her data and that it be deleted. The German cancer registry indicated that the data were automatically pseudonymized, and claimed it was not possible to re-identify the data. Consequently, they should be considered anonymous and not as personal data according to GDPR art. 4(1). The German court concluded otherwise; that it was possible to identify the data with reasonable effort, and therefore the fact that the pseudonymization was an automated process was irrelevant in this perspective.[76]

A local judgment from the Netherlands assessed the issue of identification on the basis of MAC addresses.[77] The Dutch Data Protection Agency (DPA) had fined a Municipality for temporary collection of truncated MAC addresses in order to track Wi-Fi in the centre of the city to measure effectiveness of municipal investments. The Court noted that the Dutch DPA had based the fine on an assumption of the claimant's ability to identify natural persons on the street in the early morning when there were few people out, by matching a specific person with a mobile device on the basis of hashed and truncated MAC addresses. The Court placed the burden of investigation on the DPA to demonstrate

---

[72] *C-604/22 IAB*, para 51.
[73] *C - 319/22 Scania*.
[74] *C - 319/22 Scania*, para 46-8.
[75] 21 K 1802/21 *VG Hamburg* [2021] ECLI:DE:VGHH:2022:072821K18022100 The Administrative Court of Hamburg.
[76] *21 K 1802/21 VG Hamburg* , para 75.
[77] ZWO 22/775 *Rb. Overijssel* [2924] ECLI:NL:RBOVE:2024:594 Overijssel District Court

that identification was reasonably likely and found that the DPA had not sufficiently investigated whether these methods actually make it possible to determine the identity of a user of a mobile device 'with the naked eye'.[78] The judgement states that the DPA should have investigated the identification risk considering the costs, time and available resources required for identification given technology today and technological developments.

## 4.4 Relevant guidelines and opinions on anonymization and pseudonymization

There are several guidelines and opinions available from the EDPS and EDPB on the topics of anonymization and pseudonymization in the context of AI, yet none specifically address the case of synthetic data.

The legacy WP29 group has issued two opinions that are particularly relevant to the classification of synthetic data as personal data or anonymous data, one on the concept of personal data (2007) and the other on anonymization techniques (2014).[79,80] The opinions from the legacy WP29 may have an impact on the understanding of the GDPR although both technological development and legal development through case law have evolved since their creation. The 2014 Opinion on Anonymization Techniques provides a narrow understanding of the wording in the DPD.[81] They introduce three tests for evaluating the robustness of an anonymization process: Singling out, Linkability and Inference.[82] In the opinion, WP29 defines anonymization as '… processing personal data in order to irreversibly prevent identification'.[83] The term irreversible indicates that a zero risk of identification is obtainable. Since this statement, there have been significant advances in both the technical and legal fields. While technologies designed to protect individuals from identification are advancing, identification technologies have also evolved, making it increasingly difficult to guarantee anonymity. As a result, achieving irreversible anonymity has become an unrealistic standard.[84] In the same opinion, WP29 goes on to declare that ' …when a data controller does not delete the original (identifiable) data at event-level, and the data controller hands over part of this dataset (for example after removal or masking of identifiable data), the resulting dataset is still personal data.' Healthcare providers are commonly required to archive the original datasets in the electronic health records due to an obligation of treatment documentation. Health data is stored in national registries, through research and population studies. This information will not be deleted.[85] Being special categories of personal data according to GDPR art. 9, these records are protected with technical and organizational measures, and therefore not freely accessible for processing. However, in the legacy WP29 view, the result of the mere existence of this information is that health data can never be anonymized, no matter how low the risk of identification is in practical terms.

---

[78] *ZWO 22/775 Overijssel* , paras 12-3.
[79] WP29, *Opinion 05/2014 on Anonymisation Techniques* (2014)
[80] WP29, *Opinion 4/2007 on the concept of personal data* (2007)
[81] 95/46/EC Data Protection Directive (DPD)
[82] *'Singling out*, which corresponds to the possibility to isolate some or all records which identify an individual in the dataset; *Linkability*, which is the ability to link, at least, two records concerning the same data subject or a group of data subjects (either in the same database or in two different databases). If an attacker can establish (e.g., by means of correlation analysis) that two records are assigned to a same group of individuals but cannot single out individuals in this group, the technique provides resistance against "singling out" but not against linkability; *Inference*, which is the possibility to deduce, with significant probability, the value of an attribute from the values of a set of other attributes.' WP29 *Opinion 05/2014 on Anonymisation Techniques* (2014) (n 79).
[83] WP29 *Opinion 05/2014 on Anonymisation Techniques* (2014) (n 79).
[84] Paul Ohm, 'Broken promises of privacy: Responding to the surprising failure of anonymization' 57 UCLA l Rev 1701, 2009.

The WP *Opinion on the concept of personal data* includes a clue on the timespan to consider in technological developments. They apply an example where the lifetime of the data processing is the timeline to consider.[86] Should a dataset be intended to be processed for 10 years, the horizon to evaluate technological developments must be the same. This guideline envisions an adaptable system with technological and organizational measures to monitor and re-define the data, should new technology become available that will influence what is reasonably likely to be identifiable.

WP29 was replaced by EDPB in 2019. The EDPB Opinion 28/2024 was issued in December 2024, and focuses on certain data protection aspects related to the processing of personal data in the context of AI models.[87] The document provides non-exhaustive guidance on how to interpret GDPR provisions when training and deploying AI models, and states that a model can only be anonymous when the risk of identification is negligible.[88] It refers to the fact that the context in which an AI model is released – whether it is openly published or it is within a controlled environment – affects the likelihood of identification. This provides an important distinction between information that is publicly available and information that is not publicly available.

This reinforces the concept of context controls that help mitigate the likelihood of data breaches or attacks. Importantly, the guidelines do not specify that training data must be destroyed in order to claim that a model is anonymized. This shows that the legal interpretation of the concept has evolved since the WP29 and is directly relevant to anonymity of synthetic data, where the original dataset still exists in a secured context, separate from the dataset.

In the most recent *Guidelines 01/2025 on Pseudonymisation* (Adopted on 16 January 2025), EDPB aims to clarify the concept of pseudonymization as defined by the GDPR. In the executive summary, the guidelines state that pseudonymized data is considered personal data when it can be linked to a natural person through additional information, <u>even if that information is not held by the same person</u>.[89] This definition seems to closely aligns with the WP29's interpretation of pseudonymized data where the mere existence of additional data – a key- means that the data can never be anonymous. Contrary to this, the main section of the guideline offers a different perspective.[90] It states that data should be considered personal if pseudonymized data can be combined with additional information by the data controller or another party using 'reasonable means.' This definition suggests that pseudonymized data can still be deemed personal data, but it does not necessarily have to be. A critical aspect here is the risk assessment. Notably, the guidelines do not explicitly state that pseudonymized data cannot be considered anonymous if the additional data that will make identification possible exists. In other words, there is no clear rejection of a risk-based approach to anonymization.

---

[85] Anne Kjersti Befring, 'Legal perspectives on the patient record (original title: Rettslige perspektiver på pasientjournalen)' 20 Michael 97, 2023.
[86] *WP29 Opinion 4/2007 on the concept of personal data* (2007).
[87] EDPB, *EDPB Opinion 28/2024 on certain data protection aspects related to the processing of personal data in the context of AI models* (2024).
[88] EDPB *EDPB Opinion 28/2024 on certain data protection aspects related to the processing of personal data in the context of AI models* (2024) (n 87), para 46.

Commonly referenced guidance documents from national supervisory authorities include The *Commission Nationale de l'Informatique et des Libertés (*CNIL) in France, The Irish Data Protection Commission and The Spanish AEPD together with EDPS.[91,92,93]

The concept of irreversibility from WP29 has been echoed in guidance documents from national DPAs, for example in the Irish DPA's Guidance on Anonymization and pseudonymization in 2019 or the Norwegian DPA's *Anonymization of personal data* from 2015.[94,95] These types of statements sometimes go too far compared to the direction case law is taking, in defining that anonymity must be absolute and irreversible. Examples of this have been reproduced and discussed by Befring in several articles.[96]

The guidance issued jointly by the European Data Protection Supervisor and the Agencia Espanola Protección Datos highlights that it may not always be feasible to reduce the 'probability of re-identification of a dataset to zero.' Instead, it emphasizes that a strong anonymization process seeks to lower the re-identification risk to below a specific threshold, rather than guaranteeing 100 % anonymization.[97]

---

[89] EDPB, *EDPB Guidelines 01/2025 on Pseudonymisation Adopted on 16 January 2025 (Adopted - version for public consultation)* (2025), executive summary.
[90] EDPB *EDPB Guidelines 01/2025 on Pseudonymisation Adopted on 16 January 2025 (Adopted - version for public consultation)* (2025), para 22.
[91] CNIL, 'Scientific research (excluding health): Issues and advantages of anonymization and pseudonymization. Original title: Recherche scientifique (hors santé) : enjeux et avantages de l'anonymisation et de la pseudonymisation' (2022) <https://www.cnil.fr/fr/recherche-scientifique-hors-sante-enjeux-et-avantages-de-lanonymisation-et-de-la-pseudonymisation> accessed 7.3.2025.
[92] The Irish Data Protection Commission (DPC), *Guidance on Anonymisation and Pseudonymisation* (2019).
[93] AEPD and EDPS, *10 misunderstandings related to anonymisation* (2021).
[94] *Guidance on Anonymisation and Pseudonymisation, Data Protection Commission Ireland (2019)* , page 2.
[95] The Norwegian Data Protection Agency, *Anonymization of personal data - Guidance* (2015), page 6.
[96] Anne Kjersti Befring, 'Norwegian Biobanks: Increased Complexity with GDPR and National Law.' in *GDPR and Biobanking Individual Rights, Public Interest and Research Regulation across Europe* (Springer Publishing Company 2020), p. 323-44.; Befring, *Health law (original title: Helseretten)*, chapter 10.

The value of the data will influence motivation for an identification attack. In its 'Guidance on Anonymisation and Pseudonymisation', the Irish Data Protection Commission suggested that organizations should consider the sensitivity of the personal data, as well as its value.[98] Although the UK is no longer in the EU, some guidance resources may still be relevant. The UK Information Commissioner's Office (ICO) introduces the concept of a *motivated intruder*.[99] The more motivated a potential intruder will be to identify data subjects, the more an organization should expect extreme measures to be used for identification. The motivated intruder could be anyone who gets access to the information and should be assumed to be reasonably competent. This represents a more cautious approach than the concept of an *average reader* that was suggested and refuted in OC v EC.[100] The 2025 updated ICO guidelines on anonymization incorporate the idea of identification risk being a spectrum from high to low risk, with a blurred band where the risk must be assessed based on context specific factors.[101]

## 4.5   Short overview of the relevant human rights and principles

EU secondary law is built on and harmonized with the EU charters and international human rights conventions and The primary objective of the GDPR is to safeguard the fundamental rights and freedoms of natural persons, with recital 2 highlighting the right to personal data protection.  Recital 4 emphasises that processing of personal data should be designed to serve mankind, and that data protection is not an absolute right- it must be considered in relation to its function in society and be balanced against other fundamental rights.[102]  Similarly, The Council of Europe's modernized Convention 108+ on data protection is closely related to the GDPR and states in its preamble that 'the right to the protection of personal data must be assessed in relation to its role in society and must be reconciled with other human rights and fundamental freedoms.'[103]

Human rights can be both negative rights – something the individual should be protected from, and positive rights – proactive measures that society should provide to individuals. Both protections and

---

[97] AEPD and EDPS *10 misunderstandings related to anonymisation* (2021) (n 93).
[98] The Irish Data Protection Commission (DPC) *Guidance on Anonymisation and Pseudonymisation* (2019) (n 92).
[99] Information Commissioner's Office (ICO), *Anonymisation: Managing Data Protection Risk Code of Practice* (2012).
[100] *C- 479/22 OC v Commission*.
[101] UK Information Commissioner's Office (ICO), 'ICO guidance on Anonymisation' (*ICO*, 2025) <https://ico.org.uk/for-organisations/uk-gdpr-guidance-and-resources/data-sharing/anonymisation/> accessed 3. April 2025
[102] Regina Becker et al., 'Purpose definition as a crucial step for determining the legal basis under the GDPR: implications for scientific research' 11 Journal of Law and the Biosciences lsae001, 2024.
[103] Convention 108 + (n 26).

Processing of synthetic data in AI development for healthcare and the definition of personal data in EU law rights to proactive measures must be considered. When categorizing and processing synthetic data, the rules on the rights to privacy and data protection in the EU Charter articles 7 and 8 and the right to respect for private and family life as outlined in art. 8 of the European Convention on Human Rights (ECHR) can serve as a foundation for the safeguarding of personal data.[104] The terms data protection and privacy are often used interchangeably, yet the two concepts are not identical. Data protection concerns only information about a person, not the physical person themselves. Protection of privacy on the other hand, originates in the physical sphere. It concerns protecting the physical person, family and home. This extends beyond data and encompasses both the physical person and protection of data in the digital spaces the citizens occupy. In the synthetic data literature, the term privacy risk is commonly used synonymously with data protection.

Interference or limitations to the rights to data protection and privacy may be justified by the right to life in the EU Charter art. 2 (1) and ECHR art. 2.[105] Access to medical treatment methods can be crucial to safeguarding this right. Additionally, the right to health is explicitly established in the EU Charter art. 35 and the ICESCR art. 12.[106] Tied to development of new tools and technology that may benefit health, the right to benefit from scientific and technological development found in UDHR art. 27, ICESCR art. 15 and the EU Charter art. 13 may be relevant when synthetic data is used for the development of knowledge, methods or tools with the goal to improve healthcare.[107] Similarly, art. 15 of the Oviedo Convention stresses the 'need for international cooperation so that all humanity may enjoy the benefits of biology and medicine', accentuating the importance of data sharing to the progress in medicine and healthcare.

The values that the EU is founded on as stated in TEU art. 2 include equality and non-discrimination specifically. The safeguarding of equal treatment as well as the protection against discrimination in the Charter art. 21 may have implications for how personal data is processed and for the development of tools for healthcare delivery, in how bias in the dataset is managed.[108] The principle of non-discrimination is rooted in the idea that all individuals are entitled to equal treatment and access to opportunities without unfair distinctions based on characteristics such as gender, race, religion or disabilities.

---

[104] The right to Privacy is also enshrined in the UN UDHR art. 12 and the Oviedo Convention art. 10. The right to data protection is also enshrined in TFEU art. 16, TEU art. 39 and in Convention 108.
[105] The right to life is also found in the UN UDHR art. 3, the UN Child convention art. 6.
[106] The right to healthcare is also found in the UN Child convention art. 24 and the Oviedo Convention art. 3 by the Council of Europe.
[107] The right to benefit from free sharing of scientific progress and technological development is also established in TFEU art. 179 (ex-art. 163 TEC) and TEU art. 3.3.
[108] The right to non-discrimination is also found in UDHR art. 7, ICESCR art. 2-3, ECHR art. 14 the Oviedo convention art. 11 and the UN child convention art. 2.

Processing of synthetic data in AI development for healthcare and the definition of personal data in EU law

For rights other than the right to life, proportionality assessments are relevant to assessing the risks and benefits associated with sharing medical knowledge through synthetic data generation for the development of AI.

The principles relating to processing of personal data in art. 5 of the GDPR are aligned with those of health law and human rights, emphasizing the values of legality, fairness and transparency. These principles are intricately linked to individual autonomy and the prohibition of discrimination, without compromising the fundamental tenets of health law. The same holds true for the principles of purpose limitation, data accuracy and storage limitation. The principle of confidentiality is intrinsically linked to the right to privacy, recognized as a fundamental human right. The principle of data minimization may, however, be at odds with the need for large amounts of data to develop AI models and with health law principles of due diligence.

Table 1 outlines the human rights of international treaties and EU primary law instruments that are relevant to development of AI in healthcare: the rights to data protection and privacy, the rights to life, to health and to healthcare, the right to non-discrimination and the right to benefit from science and technological development.

# Processing of synthetic data in AI development for healthcare and the definition of personal data in EU law

*Table 1 Human rights relevant for the use of synthetic data in AI applications for healthcare delivery*

| | UN | | | The Council of Europe | | | EU | | |
|---|---|---|---|---|---|---|---|---|---|
| | UDHR | UN Child convention | ICESCR | ECHR | Oviedo convention | Convention 108+ | EU Charter | TFEU | TEU |
| The right to life | Article 3 | Article 6 | | Article 2 Right to life | | | Article 2 Right to life | | |
| The right to healthcare | | Article 24 | Article 12 | | Article 3 Equitable access to healthcare. | | Article 35 Healthcare | | |
| The right to non-discrimination | Article 7 | Article 2 | Article 2 Article 3 | Article 14 Prohibition of discrimination | Article 11 Non-discrimination | | Article 21 Non-discrimination | | Article 2 |
| The right to access to science & technological development | Article 27 | | Article 15 | | Article 15 | | Article 13 Freedom of the arts and sciences | Article 179 (ex-Article 163 TEC) | Article 3.3 |
| The right to privacy | Article 12 | | | Article 8 Right to respect for private and family life | Article 10 Private life and right to information | | Article 7 Respect for private and family life | | |
| The right to data protection | | | | | | All | Article 8 Protection of personal data | Article 16 (ex-Article 286 TEC) 1. | Article 39 |

# 5 The legal provisions and perspectives on sharing synthetic data

## 5.1 Legal dogmatic analysis

### 5.1.1 The definition of personal data

The definition of personal data in GDPR art. 4(1) and the concept of identifiability, elaborated in recital 26, is closely linked with the existence of additional information that can aid in identification, or pseudonymous data as defined in GDPR art. 4(5). In this context, the discussion concerns first whether a synthetized dataset can be defined as pseudonymous, and when the existence of additional information influences whether it is considered within the scope of the GDPR.

*Real data-inspired* synthetic data, which has been generated by learning the statistical properties of a real dataset, differ from the traditional pseudonymous data in that there is no one-to-one key for re-identification. It can be argued that certain additional information such as knowledge of the model, the hyperparameters used, a similar dataset to the training dataset or even the original training dataset can improve the probability of a successful inference attack and that synthetic datasets may therefore be considered pseudonymous.[109]

The WP29 view on the *opinion on anonymization techniques* was that the mere existence of the original dataset that was pseudonymized automatically precludes it from being anonymous since the DPD recital 26 calls upon assessing 'all the means reasonably likely to be used (…) by the controller or any other person'.[110,111] The WP29 view leads to a conclusion that health data, when retrieved from electronic health records, should not be considered anonymous for as long as the patients are alive. Despite still being present in national guidelines that have adopted this view, the idea that a dataset can never be anonymous if the original data exists has been contradicted by CJEU through cases like Breyer, IAB and Banco Popular. In Breyer, the question of reasonable means was evaluated from the perspective of whether it was likely that one party could combine the two datasets to allow identification. Breyer was based on the legacy DPD, where the phrase in recital 26 regarding identifiability was 'any other person'. Kuner, Bygrave and others point to a possible narrowing of the scope in GDPR recital 26 from recital 26 DPD in the changing the of this phrase to 'another person', to denote who may constitute a legally relevant agent of identification, in addition to the controller.[112] This can also be interpreted as a clarification that the term refers to the perspective of the specific agent that should be considered.

---

[109] PHG *Are synthetic health data 'personal data', PHG foundation* (2023) (n 12).
[110] WP29 *Opinion 05/2014 on Anonymisation Techniques* (2014) (n 79).
[111] The phrasing "any other person" has been changed to "another person" in GDPR.
[112] Kuner et al., *The EU general data protection regulation: a commentary* (n 51).

Processing of synthetic data in AI development for healthcare and the definition of personal data in EU law

In the Banco Popular case and supported by the Advocate General Opinion on the appeal case C-413/23 P[113], the assessment of the identification risk is predicated on the likelihood that the data recipient has the requisite means to gain access to the additional information that may facilitate identification. Specifically, the ruling implies that the same information may qualify as personal data for one party but at the same time not for another, depending on the practical possibility of re-identification at the hands of the recipient. In the Banco Popular appeal, the Advocate General specifies in his opinion that the data should not be considered personal in the hands of Deloitte, yet it must in the hands of SRB since they are in possession of the identification key.[114] A data controller possessing additional information that enables re-identification will therefore face processing requirements, as the dataset will fall within the material scope of GDPR. Conversely, for other parties lacking the means to access such additional information, the pseudonymized dataset becomes anonymous in their hands and has no restrictions on processing.

The concept of dynamic personal data is not without controversy and its potential to create divergent processing requirements for the same dataset across different stakeholders has been the subject of ongoing academic discourse.[115] Rubinstein and Hartzog categorizes the sides in the technical anonymization debate as 'formalists' and 'pragmatists', where formalists advocate stringent definitions that are general, and pragmatists argue that one must consider the context and existing risk-reducing measures.[116] Advocate General M. Campos Sánchez-Bordona, in his opinion in the Breyer case, also deliberated on the opposing academic fronts in the definition of personal data.[117] He made a distinction between those wanting an 'objective' or 'absolute' criterion for defining identifiability – similar to the formalists, and those favouring 'subjective' criteria – similar to pragmatists.

An argument for the objective or formal approach is that the consistency of a homogenous approach will make it easier to apply, and that there is a need for a precautionary considerations, as technological advancements and access to more sophisticated tools are expected to make identification easier.

Although these are valid concerns, Advocate General M. Campos Sánchez-Bordona warns against letting the precautionary principle cloud the intention of the formulations in the regulation and that the interpretation of recital 26 of DND would be *'"the means likely reasonably to be used" by <u>certain</u>*

---

[113] *2025 Opinion of Advocate General Spielmann regarding case Case C-413/23 P European Data Protection Supervisor v Single Resolution Board (Appeal case of the Banco Popular case)*, paras 43-4.
[114] Opinion of Advocate general Spielmann, C-413/23 P, para 69.
[115] Alexandre Lodie, 'Are personal data always personal? Case T-557/20 SRB v. EDPS or when the qualification of data depends on who holds them', 2023 .
[116] Ira S Rubinstein and Woodrow Hartzog, 'Anonymization and risk' 91 Wash L Rev 703, 2016.
[117] *Opinion of Advocate General M. Campos Sánchez-Bordona regarding Case C-582/14 Patrick Breyer v Bundesrepublik Deutschland* [2016] ECLI:EU:C:2016:339 , para 52.

*third parties'*.[118] He stresses the importance of using the recital 26 criteria of effort needed for identification to be able to distinguish between likely and unlikely means and argues that if not, it would always be possible to imagine that someone now or in the future would have additional data that could assist in the identification of a user – no matter how inaccessible it would be to the data holder.[119] The result would be an unproportionally broad definition of personal data, the regulation becoming as Purtova warned 'the law of everything'.[120] Among others, El Emam and Alvarez had criticised WP29 for using the term 'irreversible' anonymization as the recommended techniques to achieve this had not been shown to work widely in practice.[121]

The EDPB opinion on personal data in the context of AI refers to the fact that the context in which an AI model is released – whether openly without control over possible identification attempts that may be made, or whether within a controlled environment – affects the likelihood of identification.[122] This may suggest that data segmentation controls, when managed by a single controller, could effectively shield identifying data from non-identifying data, allowing for a more secure separation of sensitive information.[123] The term de-facto anonymization is used to describe data in a context where identification is considered practically impossible due to controls in place, while a theoretical possibility of identification may still remain.[124]

There were great expectations to the EDPB Guideline 01/2025 on pseudonymization that they would go beyond existing guidance in their interpretation of the legal criteria of pseudonymization as defined in art. 4(5) of the GDPR with recitals 26 and 28.[125] The first section of the Guidelines aims to interpret the concept of pseudonymization, yet the recent CJEU cases are not referenced. While the guideline does not explicitly state that pseudonymized data can never be considered anonymized, it also does not shed much light on the practical interpretations of pseudonymization and so does not bring practitioners much closer to an understanding of practical implications in the matter of synthetic data than they will get by reading the actual legal text. Still, although they can be criticized for not being dynamic enough in the context of the benefits that new technologies can have in the health field, the guideline marks as a shift from WP29 towards a more flexible interpretation of personal data.[126] This shift is also evident when the earlier EDPS/AEPD guidance '10 misunderstandings related to anonymization' uses the term 'degree of anonymization' and emphasizes there will often be a residual identification risk.[127]

---

[118] Opinion of Advocate General M. Campos Sánchez-Bordona, C-582/14, para 65-8.
[119] Opinion of Advocate General M. Campos Sánchez-Bordona, C-582/14, para 68.
[120] Nadezhda Purtova, 'The law of everything. Broad concept of personal data and future of EU data protection law' 10 Law, Innovation and Technology 40, 2018.
[121] Khaled El Emam and Cecilia Alvarez, 'A critical appraisal of the Article 29 Working Party Opinion 05/2014 on data anonymization techniques' 5 International Data Privacy Law 73, 2015.
[122] EDPB *EDPB Opinion 28/2024 on certain data protection aspects related to the processing of personal data in the context of AI models* (2024).
[123] Sophie Stalla-Bourdillon, 'Identifiability, as a Data Risk: Is a Uniform Approach to Anonymisation About to Emerge in the EU?' Available at SSRN, 2025.
[124] Bundesverband der Deutschen Industrie e.V., *Anonymization of personal data* (2021).
[125] Sophie Stalla-Bourdillon, *The EDPB 01/2025 Guidelines on Pseudonymisation: A Step in the Right Direction?* (European Law Blog, 2025).

Case law has a higher relevance than soft law and it becomes clear from the analysed cases that the 'identifiability criteria from GDPR recital 26', as they are coined by Finck and Pallas, must always be applied to assess the risk of identification, irrespective of whether the data is pseudonymous or not.[128] Although establishing a dynamic definition of personal data, the CJEU maintains a broad definition where almost anything, given the right context, could become personal data.[129] Similarly, the risk assessment is applied strictly. The Advocate General Opinion on the appeal case C-413/23 P, stated that it is only when the risk of identification is 'non-existent or insignificant that data can legally escape classification as 'personal data'.'[130]

For the case of synthetic data and inference attacks it will be relevant to discuss the opportunity for identification from the perspective of the attacker and what additional information they may have access to. It is not decisive whether synthetic data is considered pseudonymous or not, since pseudonymous information can be considered anonymous according to the criteria for identifiability. The core discussion is *what constitutes a reasonable likelihood and proportionality as long as the GDPR alone is the primary regulation for the processing of synthetic data*. It is necessary to assess first whether the agent with access to the synthetic data is also likely to have access to additional information necessary for identification and then the likelihood that the agent could identify natural persons on this basis. The empirical case of synthetic data below incorporates this dynamic definition of personal data.

### 5.1.2 Assessment of synthetic data and identifiability

Identifiability is key to the definition of whether synthetic data is personal data according to the definition in GDPR art. 4(1). The risk-based evaluation of identifiability involves assessing *all objective factors* like cost and time required given available technology and technological developments. The case OC v Commission emphasizes that the effort needed must be balanced against likely motivation based on the practical opportunity or means to access the information necessary to identify a natural person and combine this to make identification possible.[131] Motivation for identification from synthetic data can be based on both the value of the original data and the success one can expect from an identification attack. Finck and Pallas introduce a flowchart for assessments to determine whether a dataset is defined as personal data which helps clarify which aspects to evaluate. [132] The flowchart builds on considerations from recital 26 regarding the reasonable likelihood to identify a person: cost, time and technological developments. It expands with elements from relevant case law: motivation, competence and available resources at the hands of the attacker.  Available additional information should be taken into account, as well as potential application of advanced methods for identification. The model is in line with the conclusions of the legal analysis in this article. Still, for practical application in the case of synthetic data, there is a need to explore how to weigh them against each other to decide what constitutes a reasonably likely and what constitutes a reasonably unlikely scenario.

---

[126] Legido-Quigley et al., 'Data sharing restrictions are hampering precision health in the European Union', 2025(n 4).
[127] AEPD and EDPS *10 misunderstandings related to anonymisation* (2021) (n 93).
[128] Michèle Finck and Frank Pallas, 'They who must not be identified—distinguishing personal from non-personal data under the GDPR' 10 International Data Privacy Law 11, 2020.
[129] Purtova, 'The law of everything. Broad concept of personal data and future of EU data protection law', 2018 (n 120).
[130] Opinion of Advocate General Spielmann, C-413/23 P, para 57.

Considerations of technological developments are central to the definition of personal data, yet the timeline one must consider is not clear. The academic discourse focuses on whether one should consider only expected new developments within existing technologies or extend the analysis to the possible effect of current uncertain research explorations or unexpected 'black swan events' such as quantum computing 'breaking the internet'.[133] In The Code Book, Simon Singh describes the cat and mouse games of encryption to safeguard information against cryptanalysts that become increasingly relevant in our current age of Information.[134] New technologies may increase the risk to data confidentiality, but also introduce new methods for protection in the ongoing arms race between de-identification and re-identification techniques.[135] Technological uncertainty highlights the need for a dynamic approach to defining risk. WP29 advice that one could consider the planned timeline of the processing of the data may be relevant to consider.[136] This approach is challenging due to the unpredictable nature of technological developments, which can both strengthen or weaken the protection of the data. Technological advancements may reduce the effort required for identification over time. The sensitivity and value of health data is assumed to be relatively high for contemporary data but can also become less relevant over time (see Figure 8).

Since GDPR does not apply to deceased persons, the expected lifetime of a given disease in the cohort can be particularly relevant in healthcare unless hereditary information could risk confidentiality breaches on for living relatives. Identification risk after this timeline is irrelevant to their possible classification as personal data.

---

[131] *C- 479/22 OC v Commission*
[132] Finck and Pallas, 'They who must not be identified—distinguishing personal from non-personal data under the GDPR', 2020 (n 128).
[133] Finck and Pallas, 'They who must not be identified—distinguishing personal from non-personal data under the GDPR', 2020 (n 128).
[134] Simon Singh, *The code book*, vol 7 (Doubleday New York 1999).
[135] Bellovin, Dutta and Reitinger, 'Privacy and synthetic datasets', 2019
[136] WP29 *Opinion 4/2007 on the concept of personal data* (2007) (n 86).

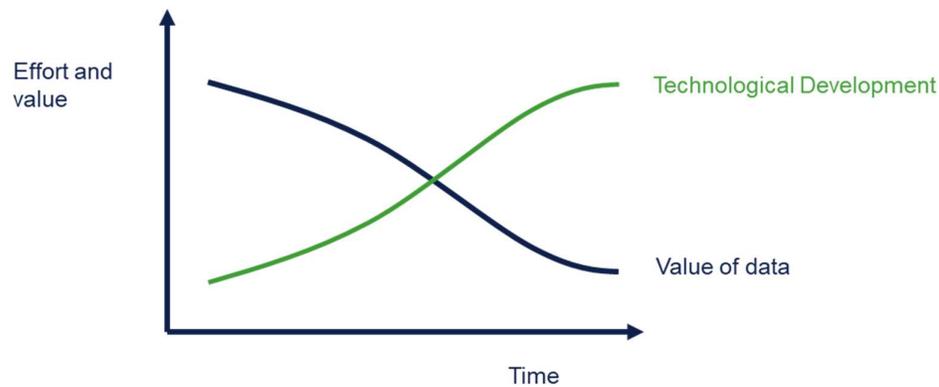

*Figure 8 Technological developments can decrease the effort of identification, while the value of health data may diminish over time, lowering the motivation for an identification attack.*

*Alt text: A graph with time on the x-axis and opportunity and value on the y-axis shows two lines. One – the value of data, starts high but decreases over time. The other, Technological developments, starts low and increases over time.*

The likelihood of attributing synthetic data to a person depends on the actual means or opportunities and whether it is reasonably likely given the motivation and the effort and resources required. To understand how the factors of opportunity, motivation and effort should be weighted to decide what is a proportionate effort and therefore practically and reasonably likely, case law provides examples that seem to be on either side of the scale: either very clearly a disproportionate effort as in illegal activity from Breyer or very clearly a proportionate effort that must be seen as reasonably likely, as in Breyer and IAB when the information is entitled to be shared upon request or as in OC v Commission when it can be found through an internet search.

Figure 9 summarizes the criteria that should be included in a consideration of the definition of personal data. It includes an overview of the conclusions EU case law has given to shed light on their interpretation.

Processing of synthetic data in AI development for healthcare and the definition of personal data in EU law

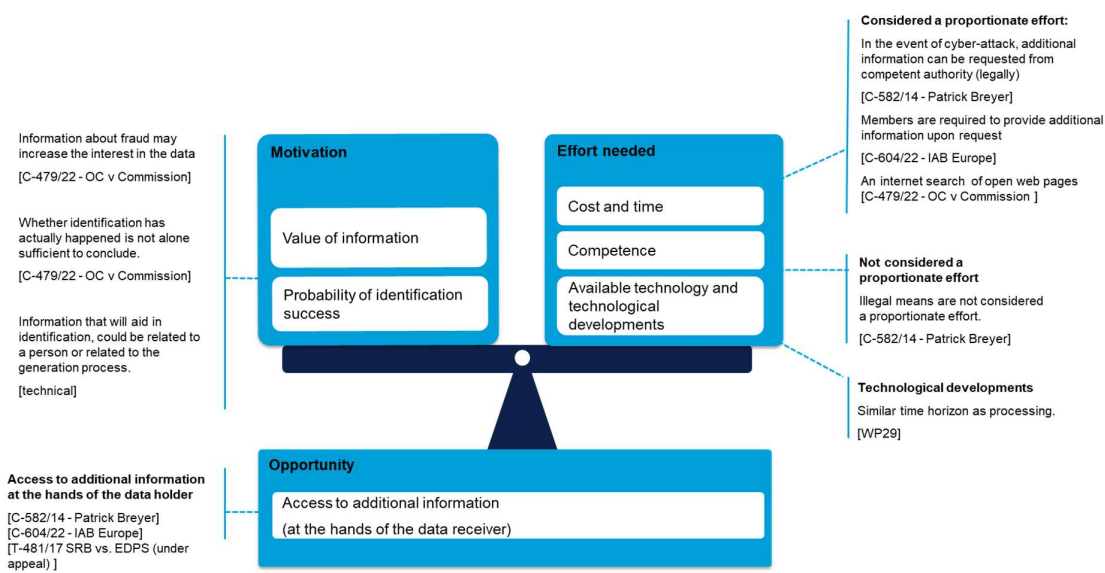

*Figure 9 Elements from the legal analysis to guide evaluation of identifiability according to GDPR. Given the practical opportunity for identification, motivation and effort are weighed against each other.*

*Alt text: The image shows a scale balancing between a box called Motivation and a box called Effort needed, standing on top of a box called Opportunity. References to relevant cases and soft law is written on the sides.*

Examples of a proportionate effort in Breyer was that in the event of a cyber-attack, additional information could be requested legally from the competent authority. In IAB and Scania, members were required to provide additional information upon request. In OC v Commission, a simple internet search was unsurprisingly deemed proportionate. These examples on either side of the scale leave a large grey zone open for interpretation. When the risk is low, the principle of proportionality and human rights, balancing the benefits versus the risks of sharing synthetic data, could be relevant to consider.

### 5.1.3 Elaboration of the significance of human rights considerations and principles

The principles and human rights that are enshrined in EU law and international conventions provide a foundation in the evaluation of whether synthetic data should be classified as personal data. Articles 7-8 of the EU Charter guarantee the rights to privacy and data protection respectively, and these must be carefully balanced against the potential benefits of data sharing, particularly in the context of healthcare and scientific advancement.[137] Key to this balancing act is the principle of proportionality, which is central to both EU law and international conventions.

---

[137] The right to Privacy: EU Charter art. 7, UDHR art. 12, ECHR art. 8, Oviedo art. 10.
The right to data protection: EU Charter art. 8, Convention 108+ and TFEU art. 16.
The right to health: UN child convention art. 24, ICESCR art. 12, Oviedo art. 3, EU Charter art. 35.
The right to life: UDHR art. 3, UN child convention art. 6, ECHR art. 2, EU Charter art. 2.

Processing of synthetic data in AI development for healthcare and the definition of personal data in EU law

The European Court of Human Rights has repeatedly held that the European Convention on Human Rights is a 'living instrument' subject to dynamic interpretation to reflect societal changes and technological advances. The rapid development of complex technologies necessitates dynamic legal development, where fundamental principles and human rights are a foundation. In this context, Befring highlights that international law provides a framework for navigating rights and assessment issues in scenarios where technologies have the potential to impact life, health and the future of humanity.[138]

Under art. 52 of the EU Charter and art. 69 of the Treaty on European Union (TEU), proportionality assessments require weighing the risks and benefits of any interference with fundamental rights. Similarly, ECHR art. 8(2) calls for a proportionality test when restrictions are imposed on privacy rights. Restrictions on the right to privacy can be imposed based on exceptions for- among others- the protection of health, when there is a legitimate objective that follows from the law.[139] This principle aligns with the framework established in EU law, which also serves as the basis for assessments of proportionality.

In several rulings, the European Court of Human Rights has balanced the interests of society against those of the individual. In some cases, the common good such as public health has taken precedence over individual rights. A notable example of this is the case *Vavřička and Others v. the Czech Republic* (2021), where the European Court of Human Rights held that a requirement for children to be vaccinated in order to attend preschool was proportionate and legitimate under art. 8(2) of the ECHR.[140] The Court acknowledged the importance of protecting public health, stating that 'in the context of vaccination, the need to protect the rights of others and the public interest outweighed the applicants' rights to private life'.

In the context of synthetic data, applying restrictions on its processing could undermine its primary advantage—enabling sharing and lowering the cost of data. Lowering the administrative burden is a prerequisite to facilitating sharing. While synthetic data reduces privacy risk compared to real data, it is not necessarily devoid of residual risk. [141]  Privacy risk can be managed through specific generation

---

  The right to access to scientific and technological developments: UDHR art. 27, ICESCR art. 15, Oviedo art. 15, EU Charter art. 13, TFEU art. 179, TEU art. 3.3.
  The right to non-discrimination: UDHR art. 7, UN child convention art. 2, ICESCR art. 2-3, ECHR art. 14, Oviedo art. 11, EU Charter art. 21, TEU art. 2.
[138] Anne Kjersti Befring, 'Transformation of Medical Care through Gene Therapy and Human Rights to Life and Health–Balancing Risks and Benefits' 29 European Journal of Health Law 359, 2022.
[139] See also C-249/96 *Lisa Jacqueline Grant v South-West Trains Ltd* [1998] ECR I-621 CJEU, para 44 which refers to proportionality considerations of ECHR art. 8(2).
[140] *Vavricka and others v. the Czech Republic* [2021, Applications nos 47621/13 and 5 others] European Court of Human Rights .
[141] Jean-Francois Rajotte et al., 'Synthetic data as an enabler for machine learning applications in medicine' 25 Iscience, 2022; Mark Elliot et al., 'The anonymisation decision-making framework', 2016.

methods. Unfortunately, measures that reduce privacy risks also tend to lower the utility of the data, in turn introducing safety risks that might counteract the benefits to public health and well-being.[142,143]

A key concern in the use of synthetic data is its representativity – its ability to properly reflect the population for which it is intended. To produce meaningful and actionable data, synthetic datasets must capture the full spectrum of population diversity, including minority groups. Failure to account for this could lead to amplified biases, which may undermine the precision of AI models and could result in adverse outcomes, especially for marginalized groups.[144] If executed consciously, the generation of synthetic data has been shown to be a method of adjusting for inherent biases in collected data, as well as generating datapoints to replace missing data in existing datasets to improve the structural quality.[145] The use of synthetic datasets can thus support the right to non-discrimination in art. 21 of the Charter and art. 2 of the TEU and art. 14 of the ECHR. In high-risk applications like healthcare, AI models must ensure that they are fair and equitable across diverse demographic groups.[146]

Several considerations must be made when balancing human rights in the context of using synthetic data for AI development for healthcare.

*First,* the responsible use of synthetic data to develop AI tools in healthcare requires quality assurance and a balancing of all relevant quality dimensions – representativeness, privacy, fairness and carbon footprint.[147] Another significant consideration is the balance between transparency and confidentiality. While transparency in AI model development can increase trust and ensure equal treatment[148], it also raises the risk of exposing sensitive information. Detailed insights into the data used to train AI models could lead to inferences that may compromise confidentiality. This trade-off underscores the importance of properly managing synthetic data generation to minimize risks while maintaining its utility for AI development.

For the proportionality assessments, considerations of how synthetic data can mitigate risks to representativity, privacy or bias and simultaneously maximize the benefits of AI in healthcare development must be reflected.

---

[142] Debolina Ghatak and Kouichi Sakurai, 'A Survey on Privacy Preserving Synthetic Data Generation and a Discussion on a Privacy-Utility Trade-off Problem' (International Conference on Science of Cyber Security, 2022).
[143] Matthew Fredrikson et al., 'Privacy in pharmacogenetics: An {End-to-End} case study of personalized warfarin dosing' (23rd USENIX security symposium (USENIX Security 14), 2014)
[144] Karan Bhanot et al., 'The problem of fairness in synthetic healthcare data' 23 Entropy 1165, 2021.
[145] Zhenchen Wang et al., 'High-fidelity synthetic data applications for data augmentation', 2024
[146] UDHR art. 7, ECHR art. 14, ICESCR art. 2 and 3, Oviedo Convention art. 11, EU Charter art. 21, TEU art. 2.
[147] Vibeke Binz Vallevik et al., 'Can I trust my fake data – A comprehensive quality assessment framework for synthetic tabular data in healthcare' 185 International Journal of Medical Informatics, 2024.
[148] Requirements for transparency can be found in AI act AI Act , art. 13.

Processing of synthetic data in AI development for healthcare and the definition of personal data in EU law

*Secondly, o*ne of the principal human rights considerations is the right to health, as enshrined in art. 35 of the EU Charter, which guarantees the right to access to preventive healthcare and to benefit from medical treatment under nationally established conditions. The right to health is closely linked with the right to life in the Charter art. 2. In the hierarchy of human rights, the right to life is called '*the supreme value*' and is often referred to as an absolute right. [149]

Human rights are used by the EU courts to clarify, argue for and balance different considerations, and the courts have established that the right to health in art. 35 of the EU Charter is 'one of the overriding reasons of public interest recognized by EU law.'[150] For example, in a case on establishment of para-pharmacies, it was determined that the protection of public health was an overriding concern to justify restrictions on the freedom of establishment.[151] In a case on veterinary medicine, it was held that protection of public health was an overriding requirement that justified measures to ensure reliable and good quality supply of medicinal products. [152] Art. 168 TFEU states that 'A high level of human health protection shall be ensured in the definition and implementation of all Union policies and activities.' TFEU art. 168 was used during the COVID-19 pandemic to justify the necessity of violating other fundamental rights such as freedom of movement in order to protect the population from the spread of infection.

The development of new methods and tools may be central to sustaining the level of healthcare our society enjoys today with the projected future lack of personnel due to the demographic development in European countries.[153] In the context of developing AI models for health-related purposes, this could be framed as necessary for the protection of public health, potentially justifying the use of synthetic data with a low or moderate identification risk.

In healthcare, the potential of synthetic data extends beyond merely safeguarding privacy. It can be produced on a large scale at a relatively low cost and can facilitate the development of AI-driven diagnostics and treatment tools, especially in areas with limited real-world data, such as rare

---

[149] *Streletz, Kessler and Krenz v. Germany* [2001] Applications nos 34044/96, 35532/97 and 44801/98 European Court of Human Rights , para 94.
[150] See C-159/12 to C-161/12 *Venturini and Others* [2013] ECLI:EU:C:2013:791 CJEU, para 41-2;
C- 297/16 *Colegiul Medicilor Veterinari din România (CMVRO) v Autoritatea Naţională Sanitară Veterinară şi pentru Siguranţa Alimentelor (Veterinary medicinal products case)* [2016] ECLI:EU:C:2018:141 CJEU, para 57;
C- 230/18 *PI v Landespolizeidirektion Tirol: on prostitution and Protection of public health* [2019] ECLI:EU:C:2019:383 CJEU, para 71-2.
[151] *C-159/12 to C-161/12 Venturini* , para 41
[152] *C - 297/16 CMVRO, para 57.*
[153] Gunnar Bovim et al., *Time for action - The personnel in a sustainable healthcare service. NOU 2023: 4. Original title: Tid for handling — Personellet i en bærekraftig helse- og omsorgstjeneste* (Official Norwegian Reports (NOU) 2023)

diseases.[154,155] In rare diseases, e.g. childhood cancer, there is a lack of data to develop tools that support diagnosis or treatment. Synthetic data may be a solution to the need for knowledge to achieve targeted treatment.[156]

With the provision of the right to healthcare under the conditions set by national laws and practices, expectations of the standard of healthcare will be higher in a rich state compared to a poor state. The state's duty to provide and organize healthcare implies that it should be based on medical knowledge and technology. Several human rights may apply to future patients. For rare disease patients including paediatric cancer, children's rights in articles 6 and 24 of the UN Convention on the Rights of the Child can be relevant.

*Finally,* the right to benefit from scientific progress, as articulated in art. 27 of the UDHR and art. 13 in the Charter, defines the human right to take part in scientific progress and its benefits. The EU has embraced this principle through policies that encourage the free movement of scientific knowledge and research, particularly in the healthcare sector, as illustrated through the commitment to fostering innovation and data sharing in The European Strategy for Data.[157]

The free movement of content and knowledge for the exchange of ideas and experiences, also known as the proposed 'fifth freedom', includes several key points aimed at strengthening research, innovation and data and knowledge sharing within the EU. This commitment is further reinforced by art. 13 of the EU Charter on Freedom of the Arts and Sciences, which should be interpreted in the context of art. 179 TFEU, which states that 'The Union shall aim to strengthen its scientific and technological bases by achieving a European Research Area in which researchers, scientific knowledge and technology circulate freely'. The European Strategy for Data serves as a practical illustration of the implementation of this freedom.

Synthetic data, when managed responsibly, would facilitate the rapid development of AI tools that have the potential to revolutionize healthcare, especially in areas like rare diseases and paediatric care. The considerations of relevant human rights in the use of synthetic data use must be assessed carefully and can be applied in cases of low privacy risk, balancing the risks of potential privacy infringements with the potential societal benefits. The right to health and the protection against discrimination are arguments in favour of its use. However, careful legal and ethical considerations must guide its use to ensure that fundamental rights are upheld and societal benefits maximized.

---

[154] Mauro Giuffrè and Dennis L Shung, 'Harnessing the power of synthetic data in healthcare: innovation, application, and privacy' 6 npj Digital Medicine 186, 2023
[155] Mahesh Kumar Goyal, 'Synthetic Data Revolutionizes Rare Disease Research: How Large Language Models and Generative AI are Overcoming Data Scarcity and Privacy Challenges',
[156] Gonçalo Forjaz et al., 'Making the Case for an International Childhood Cancer Data Partnership' JNCI: Journal of the National Cancer Institute djaf003, 2025
[157] European Commission *A European strategy for data COM/2020/66 final (EU data strategy)* (2020).

## 5.2 Evaluation of the empirical case based on the legal dogmatic approach

### 5.2.1 Opportunity and means

In synthetic data, identification can potentially happen either from synthetic samples that are exact matches or very similar to original samples, or from an inference attack on the synthetic dataset using shadow-modelling techniques. The investigated scenario is whether a release of the synthetic dataset from the Florence project can lead to a risk of identification by another person than the controller.

There were no exact matches in the synthetic Florence dataset. A Membership Inference Attack (MIA) simulation was performed to investigate the robustness of the dataset against inferences, based on models that are currently considered state-of-the-art.

The empirical case concludes with a relatively low identification success of the MIA with an average ROC of 0.75 given access to all possible parameters without having the original training data. It is also noteworthy that the results of the empirical case encompass target record identification attempts where the model performs not only better, but also worse than a random guess. This result shows synthetic data can offer a certain degree of protection against MIA, with the effectiveness of the attack being significantly influenced by the selection of the target record. The attacker is likely to achieve greater success against records that contain features that differentiate them from the rest of the data (i.e. statistical outliers).

While a comprehensive reassessment and refinement of the attack model is beyond the scope of this paper, the results underscore the potential practical limitations of the attack methodology.

A consequence of applying the dynamic definition of personal data is that the risk must be assessed from the attacker's perspective and based on the additional information this attacker is likely to have access to. To define a realistic attack, an analysis was done on the additional information that could be expected to be available to another person. This would be information that is often publicly available, including details that patients may share on social media – diagnoses, dates of interventions etc. (see Appendix B on the dataset). In the literature of testing synthetic data privacy, it is not uncommon to set up the attack simulation as a situation assuming the attacker has access to the original records as 'additional information' to aid identification.[158] Unsurprisingly, this greatly increases the success of an inference attack but cannot be seen as a likely scenario.

In an attack simulation, the data controller uses the true values of the original dataset to verify the performance of the inferences. In a more realistic scenario, the attacker will not have access to the original dataset and is therefore unable to verify the performance of their attack. A Membership Inference Attack for a synthetic dataset will not provide the attacker with a certain identification but will provide a qualified guess about whether or not the patient was included in the training data (see description of the empirical case). This guess may or may not be correct. Even in the event of a successful guess by the attacker, the practical implications are contingent on the attacker's ability to confirm the suspicions. With no additional knowledge to ascertain the performance of the inference, from the perspective of the attacker the prediction will have the same value as a random guess. Even though the model actually makes a correct membership prediction- unless the prediction can be confirmed, one could argue that no information has actually been leaked.

---

[158] Gadotti et al., 'Anonymization: The imperfect science of using data while preserving privacy', 2024 (n 44)44.

### 5.2.2    Motivation

When considering the risk of whether synthetic data can be indirectly linked to a natural person in a legal perspective, the motivation must be balanced against the effort of an identification attack. A high value of the data may increase motivation (see OC v Commission).[159] The Florence dataset consists of a cohort of colorectal cancer patients. The consequence of membership disclosure of an individual will imply that this individual has at some stage in their life been diagnosed with colorectal cancer. Health data, as a special category of sensitive personal data (GDPR art. 9) could be considered to increase motivation. One can argue that there are different levels of healthcare data where the most sensitive typically is thought to be mental illness, abortion or certain chronic diseases, and the less sensitive typically would be fractures or other accident-related trauma. For cancer data, the implications of the diagnosis can be severe, yet the disease is normally not socially stigmatized.

The relatively low-quality information achieved from an unverified inference attack would be expected to significantly lower the motivation of an attack.

### 5.2.3    Effort

The empirical study shows that identification attacks from synthetic data is a resource-intensive exercise. The *time* and *cost* it can take to attempt to identify individuals in a dataset proved to be substantial in the empirical test. Meeus and others similarly found that even when running a relatively small (1000 records) dataset, fully estimating the privacy risk of a dataset on HPC facilities would take up to 250 days of compute to identify individuals.[160]

### 5.2.4    Technological developments

Today, the application of membership identification attacks requires data science competence and access to high-performance computing. Over time, more easily applied tools may be made available to allow for a broader audience and reduce the effort needed. Measures to successfully identify individuals in datasets are becoming increasingly more effective and may over time increase the probability of a successful identification attack.[161] This in turn could contribute to increased motivation.

---

[159] *C- 479/22 OC v Commission*.
[160] Meeus et al., 'Achilles' heels: vulnerable record identification in synthetic data publishing' (2023) (n 48).

The timeline for discussing technological developments must be considered. The expected period for processing of the synthetic dataset should be regarded, as should the life expectancy of the patient population. If the patients are no longer alive, the GDPR no longer applies unless specific national laws do.[162] The timeline of considering technological developments should be aligned with this perspective.

### 5.2.5 Conclusion of the empirical case

In this particular empirical study, balancing opportunity, motivation and effort shows a resulting risk of identification that is significantly low. The original training data is health data and therefore of higher value in terms of sensitivity. There is an opportunity to use state of the art technology to perform inference attacks on the synthetic data. One can expect an external party to gain access to additional information from social media or other sources, and these were included in the simulation. Even with access to this, the performance of the inference guesses from the best performing samples had a ROC AUC of 0.75 which is better than a random guess at 0.5, but far from verified information. As the attacker will not have access to the true values, they are unable to verify the performance to identify the ones that are more probable to be correct.

The effort needed was several weeks of work, a need for high data science competence and access to high performance computing. While it is not an illegal effort as in the Breyer case, this level of effort for identification is incomparable to that of any examples of proportional effort from case law.

After the effort, the resulting attack performance has a high uncertainty of whether the information can be considered a leak or whether it has any higher value than a random guess. This leads to the conclusion that motivation – although the case contains health data – is expected to be low. Expected technological developments may reduce the effort needed, but over time the value and therefore motivation will expectedly diminish. The empirical study shows that the data must most likely be considered anonymous. Identification from this synthetic dataset is not necessarily impossible, but reasonably unlikely. Evaluations should be reiterated upon larger new technological developments to verify whether the conclusion still stands.

---

[161] Yves-Alexandre De Montjoye et al., 'Unique in the crowd: The privacy bounds of human mobility' 3 Scientific reports 1, 2013.

[162] For example in Norway, patient confidentiality is regulated also after death, see Norwegian Ministry of Health and Care Services, *Regulation (Norway) LOV-1999-07-02-64 The Health Personnel Act (original: Helsepersonelloven)* (1999) , §21 with §24. The scope of Regulation (Norway) LOV-2014-06-20-42 Processing of health data for healthcare. (Original : Pasientjournalloven) in §3 and in Regulation (Norway) LOV-2008-06-20-44 Health research (original:Helseforskningsloven)  §2 also specifically state that confidentiality protection is extended after death.

## 5.3 Reflections on synthetic data and legal policy

This article examines whether and when fully synthetic data created by training a generative model on a real-world dataset can be considered non-personal data and thus processed outside the material scope of the GDPR. The empirical study above shows a relatively low risk of individuals from the original training data being identified from the synthetic data and concludes it should not be considered personal data, after weighing the high effort of an identification attack with a likely low motivation resulting from the low performance of the attack. Similarly, the 2022 NeurIPS hide-and-seek contest showed a consistent low performance in the submitted attacks.[163] Different types of cases should be examined before generic conclusions can be drawn, but there are several arguments that this is likely true for most synthetic datasets that are created through training ML models on an original dataset. These build on the dynamic definition of personal data, the fact that 100 % anonymity is not obtainable, and on human rights proportionality assessments balancing the risks to data protection with the benefits synthetic data can provide to life, healthcare and partaking in technological progress.

### 5.3.1 The dynamic definition of personal data must dictate the attack model scenarios

Synthetic data bears resemblance to pseudonymized data in that identification is contingent on supplementary information. The success of the attack will increase with access to supplementary information, and with a lower number of entries or a larger proportion of anomalies, as may be found in datasets for rare disease or small population studies. The empirical experiment builds on a reasonably likely scenario of what additional information one can reasonably assume an attacker will have access to. This includes open sources and social media, but not the entire original training dataset, e.g. actual clinical patient data. Case law dictates that in assessing reasonable means one should include any additional information that the data recipient can lawfully access, is entitled to obtain upon request or can find through a relatively simple internet search. Inference attacks in the literature often assume access to the original data, which would not reflect a likely scenario.[164]

The definition of personal data must be understood as dynamic, with the key factor being the likelihood that the data can be linked to natural persons as evaluated from the perspective of the data receiver. Exact matches in the synthetic dataset can be viewed as copies or as a mere chance result of the creative generation process. According to the infinite monkey theorem, a monkey randomly hitting keys for an infinite amount of time will almost certainly produce a given text such as the entire play of William Shakespeare, as illustrated by Émile Borel in 1913.[165] The same philosophical idea can be discussed for the generation of synthetic data. An exact match to a record can be a coincidence of the synthesis, rather than an actual copy.

---

[163] James Jordon et al., 'Hide-and-seek privacy challenge: Synthetic data generation vs. patient re-identification' (NeurIPS 2020 Competition and Demonstration Track, 2021)
[164] Gadotti et al., 'Anonymization: The imperfect science of using data while preserving privacy', 2024 (n 44)44.
[165] Émile Borel, 'La mécanique statique et l'irréversibilité' 3 J Phys Theor Appl 189, 1913.

In the empirical case, there are no exact matches between the original and the synthetic data samples. Even in cases where exact matches occur, the argument of the non-verifiable predictions extends to the concept of exact matches, when the generator creates samples that are equal to real data. While it seems intuitive to remove exact matches if found, this may be counterproductive from the perspective of the 'Swiss cheese' argument. If an attacker can identify apparent 'holes' in the synthetic dataset, this will tell them that the missing sample was actually similar to an original record. Without the additional information of the original record to compare, an attacker should not be able to ascertain whether or not a sample in a fully synthetic dataset is an exact match. From this, it can be deduced that exact matches in themself do not necessarily make identification possible.

The empirical study illustrates how the success of an inference attack would be uncertain from the attacker's perspective, since there is no original data to verify the performance of the inference model. As long as the attacker cannot verify their attack performance, this raises the question of whether the information has any higher actual value than a random guess. This uncertainty in the results is assumed to lower the motivation for an attacker to perform a laborious attack such as an inference attack.

Simulations that are built on the assumption that the attacker has access to the original training data makes it possible to confirm the performance of a prediction and demonstrate a higher identifiability risk. Two hypothetical situations are commonly used to support building the MIA on this type of scenario. Having access to the original target record will indeed increase the probability of a successful MIA prediction. The first argument posits that if a medical professional has access to the original medical records, they can utilize these records to verify the efficacy of an MIA. While this is a theoretically possible scenario, it relies on the physician processing the patient data illegally and should therefore, according to Breyer, not be considered reasonable means. When a medical doctor has access to the medical records for the purpose of treating a patient, this information cannot be used legally for a different purpose such as identifying a patient in a synthetic dataset. The second hypothetical situation concerns a patient recognizing themselves in the synthetic data, having knowledge of or access to their own medical records. One can argue this situation is similarly irrelevant. The main purpose of the GDPR is to safeguard the fundamental rights and freedoms of natural persons, specifically the right to protection of personal data.[166] Its purpose is not to protect anyone from revealing their own data to themselves. The judgments of EU law concerning the use of additional information to identify individuals do not address the risk of individuals identifying themselves. For example, in the case with license plates and Scania, the question is whether a third party can combine this license plate with a name to identify a person, not whether someone can recognize their own license plate. Recital 26 of the GDPR specifically uses the phrase 'by another person' when defining identifiability risk. This suggests that the ability to identify oneself is not decisive in determining whether synthetic data is personal data or anonymous.

---

[166] GDPR, art. 2.

The more information that is available regarding the synthetization process, the more vulnerable the dataset will be to inferencing. A dilemma may arise in terms of the requirements for transparency regarding the data utilized and the AI model, as outlined in the AI Act, given that such transparency has the potential to compromise privacy. Transparency in choice of generation model and process parameters can amplify the vulnerability to attacks, thereby necessitating the acknowledgement that potential attackers possess knowledge of and access to the same generator employed in the creation of the synthetic dataset during the assessment process.

The advisory bodies EDPD and EDPS have set more conservative limits for what constitutes personal data in the case of additional data compared to the CJEU as seen in the cases analysed. Given the secondary purpose of GDPR to facilitate data sharing, it is difficult to consolidate such a conservative view with these intentions, as it restricts the free flow of knowledge.[167] The CJEU has to a greater extent assumed that the 'spirit of the law' must be assessed.[168] In the cases concerning the definition of personal data, the Court has considered risk and proportionality, which shows that the definition is more dynamic than the advisory bodies have assumed. The EU courts' dynamic interpretations of these regulations are crucial to its understanding. Despite earlier opinions of WP29, EU case law has demonstrated that the mere existence of additional information that would make identification possible is not sufficient to consider the data identifiable if there is no reasonable probability that it will be combined at the hands of the recipient of the synthetic dataset (see Breyer, IAB, Banco Popular). As a result, a synthetic dataset could be personal data for the controller who has access to the original training dataset and at the same time anonymous in the hands of an external data holder. Consequently, the boundary between personal data and anonymous data is relative and context-specific, rather than absolute. The likelihood of identification can be significantly lower in a protected environment which can provide 'de facto anonymization'.

---

[167] Becker et al., 'Purpose definition as a crucial step for determining the legal basis under the GDPR: implications for scientific research', 2024 (n 102); Emily M Weitzenboeck et al., 'The GDPR and unstructured data: is anonymization possible?' 12 International Data Privacy Law 184, 2022.
[168] C- 26/62 *Van Gend en Loos v Administratie der Belastingen* [1963] ECLI:EU:C:1963:1 CJEU, para 12.

Processing of synthetic data in AI development for healthcare and the definition of personal data in EU law### 5.3.2 A risk-based approach to anonymity

In the context of a data-driven society, the prospect of achieving full anonymity is widely considered to be virtually unattainable.[169] Constantly evolving technological developments may necessitate updated assessments, as they may change both the effort, competence needed and the success probability of an attack. Future predictions are hard to make. The advance of quantum computing is predicted to create a paradigm shift in cryptography and produce machines that could 'break the internet' by 2030-2040.[170] On the other hand, the technology can be used to improve data protection efforts and be crypto-agile in a post quantum scenario. This new technology must be seen not just as a threat but also as a new tool in the continuing cat and mouse-game of data protection.[171] The emergence of new technologies creates a layer of uncertainty and may necessitate perpetual evaluations of risks and benefits.

### 5.3.3 Proportionality assessments and balancing risks and rewards

The application of synthetic data in healthcare has the capacity to facilitate the development of artificial intelligence models across a range of domains, including research, education, knowledge and tool development for clinical care, public health and new treatment development.[172] Both the AI Act (art. 59 1 b) and the Data Governance Act (DGA)[173] refer to synthetic data as a type of non-personal data. These statements can be understood as meaning that synthetic data is largely perceived as anonymous.[174]

Still, the fact remains that the risk-based approach to anonymity with its large undefined area of what is not considered a 'proportionate effort to make identification reasonably likely', leaves an uncertainty in the classification of synthetic data as personal or non-personal data. The sanctions outlined in the GDPR, including significant fines, may drive a precautionary approach within the EU and countries with EU agreements. For example, the UK Medicines and Healthcare Products Regulatory Agency (MHRA) generated synthetic datasets to increase reasonable access to health data for training purposes and to improve algorithms.[175] The legal assessment of the residual privacy risk resulted in the synthetic dataset being subject to restrictions, which in turn creates organizational costs.

---

[169] De Montjoye et al., 'Unique in the crowd: The privacy bounds of human mobility', 2013(n 161).
[170] 'Quantum computers will break the encryption that protects the internet- Fixing things will be tricky' *The Economist* <https://www.economist.com/science-and-technology/2018/10/20/quantum-computers-will-break-the-encryption-that-protects-the-internet?utm_medium=cpc.adword.pd&utm_source=google&ppccampaignID=18151738051&ppcadID=&utm_campaign=a.22brand_pmax&utm_content=conversion.direct-response.anonymous&gad_source=1&gclid=Cj0KCQjwurS3BhCGARIsADdUH50RRO1OmElRGbRJqVRRGgedfv-u29vlO61RS3JymZTZuX9cH9gzlL8aAnzHEALw_wcB&gclsrc=aw.ds> .
[171] Said Gulyamov, 'Quantum law: navigating the legal challenges and opportunities in the age of quantum technologies' 1 Uzbek Journal of Law and Digital Policy, 2023.
[172] Aldren Gonzales, Guruprabha Guruswamy and Scott R Smith, 'Synthetic data in health care: a narrative review' 2 PLOS Digital Health e0000082, 2023.
[173] Data Governance Act
[174] Magdalena Kogut-Czarkowska and Nayana Murali, *Synthetic data – a miracle cure or a data protection headache?* (2024) ; Lorenzo Cristofaro, *Legal status of Synthetic Data* (2023) .

Limiting the sharing of synthetic data undermines one of its core purposes: to enhance access to useful data when access to the original data is restricted. Vallevik et al. discuss an approach of ethical consequentialism and duty ethics to investigate sharing of synthetic data and conclude that the threshold for acceptable privacy risk must be considered in relation to the purpose of the processing and its benefits, meaning there may be acceptance of higher risk if the purpose is to protect other fundamental rights.[176]

Human rights and proportionality assessments are crucial both in the formulation of new regulatory frameworks and in comprehending the nuances of existing regulations.[177] In circumstances where the definition of synthetic data as personal data is unclear, or where regulations result in violations of other rights, human rights can serve as a foundation for legal arguments and as a corrective.

The right to life under art. 2 of the ECHR may include the obligation of a state to ensure access to healthcare services that are based on new knowledge. The right to health in art. 12 of the ICESCR, is a fundamental part of the right to life and the understanding of a life in dignity. The universal right to health must be seen in the context of art. 15 (b) of the ICESCR on the right to enjoy the benefits of scientific progress and its applications.

The balance between data protection and sharing medical knowledge becomes clear in the following example on risk for attribute inference attacks (AIA). Like in an MIA, an AIA is performed by training a machine learning model (the inference model) on the synthetic dataset, this time to predict a sensitive variable or 'attribute'. The trained inference model is then applied to data about natural persons in order to predict the sensitive value for these persons. An example is provided in the AEPD guide to anonymization: 'if a dataset released by a medical researcher reveals that 70% of individuals above the age of 75 have a certain medical condition, this information could be inferred about an individual who is not in the dataset'.[178] While this can be viewed as a data leak, it can also be seen as using medical knowledge. A similar prediction can be made based on statistical medical information, and while an inference model may offer a higher probability than a general statistic, it remains a prediction according to the same principle. While certain cases could be defined as a leak of personal sensitive data, the example given has the characteristics of medical knowledge, which one can argue should be shared for the benefits of scientific progress and the common good. Data representing medical knowledge should be accorded significant importance in order to facilitate the enhancement of healthcare for future generations.

---

[175] 'CPRD synthetic data'  <https://www.cprd.com/synthetic-data > accessed 15.11.
[176] Vibeke Binz Vallevik et al., 'Synthetic datasets can provide the health service with better AI models Original title: Syntetiske datasett kan gi bedre KI-modeller for helsetjenesten' Tidsskrift for Den norske legeforening, 2024.
[177] Koen Lenaerts, 'Exploring the limits of the EU charter of fundamental rights' 8 European Constitutional Law Review 375, 2012; Paul P Craig, 'Proportionality and Judicial Review: A UK Historical Perspective' General Principles of Law: European and Comparative Perspectives, ed S Vogenauer and S Weatherill, 2016 .
[178] Agencia Española de Protección de Datos (AEPD), *Basic guide to anonymization by PDCP Singapore. Original title: Guía básica de anonimización Elaborada por Autoridad Nacional de Protección de Datos de Singapur (PDPC - Personal Data Protection Commission Singapore)* (2022) .

After World War II, Hannah Arendt warned of the potential dangers of balancing individual rights against the needs of society in *The Origins of Totalitarianism*.[179] This statement must be seen in light of the atrocities that occurred during the World War, yet the current international political scene may point towards a need to continue factoring this into the discussion.

The legal analysis shows the evolution of a nuanced and dynamic understanding of the definition of personal data, which proves challenging to implement in practice. The current ambiguity leads to overly cautious interpretations of what constitutes personal data, stifling the benefits of synthetic data in research and AI development for healthcare. There is an urgent need for clarifications to guide practitioners in synthetic data generation on the delineation between personal and non-personal data. Legal considerations should balance privacy protection with broader societal and human rights considerations, ensuring that the benefits of synthetic data can be realized without compromising individual rights or public welfare.

# 6 Conclusions and considerations for synthetic data regulation in a policy perspective

## 6.1 Synthetic data should rarely be considered personal data

According to the assessments of legal sources, the definition of personal data outlined in art. 4 (1) of the GDPR can be regarded as a risk regulation. The examination of legal sources reveals the absence of a clear delineation between personal data and anonymous data in circumstances where additional information that can aid identification is accessible. This ambiguity leads to uncertainty regarding the application of synthetic data and hampers the benefits of its use. The judgments of the European Court of Justice reflect that 100 % anonymity is not a realistic expectation, and that risk assessments must be made. The risk must be considered based on the context of each specific case, and crucially – the assessments of identification risk must be built upon a likely scenario of what additional information the attacker may likely access, rather than the existence of an original dataset. Patients recognizing themselves in a dataset is not a relevant scenario to include in a risk assessment, as it is not relevant to the purpose of GDPR, as demonstrated through the evaluations found in CJEU cases.

---

[179] Hannah Arendt and Nadia May, 'The origins of totalitarianism', 1958.

Human rights and legal principles support this approach and highlight the need to consider the consequences of not making information available.

The *conclusion* from this analysis is that well-generated synthetic datasets can be used and shared independently of GDPR, based on the premise that there is minimal risk of identifying individuals. This builds on the assumption that the motivation will be low, resulting from an uncertainty in the results, the inability to verify the results and the high effort needed. Certain special cases make identification risk higher, like statistical outliers or an opportunity for the attacker to verify their information.

Furthermore, the benefits of new knowledge and methods that can save lives must be emphasized, paired with the EU's ambitions of a data sharing economy. Using synthetic data instead of real data is one way to reduce the likelihood of harm, but it does not always entirely eliminate the risk. Attempts to further minimize privacy risks may reduce the utility of the data and thereby increase the risk of harm and loss of benefits for future patients. The benefits of sharing a dataset to improve research and develop new tools to help future patients may outweigh a significantly low identification risk.

The conclusion of the analysis of case law highlights a grey area concerning what should be considered reasonably likely and what constitutes a proportional effort to identify an individual. This legal ambiguity combined with GDPR's sanctions that include hefty fines, lead to a precautionary approach. Processing synthetic data as personal data with restrictions results in adding costly administrative procedures and hinders its use as a catalyst for the development of new treatments while ensuring the confidentiality of individual patients.

The analysis points to that acceptable privacy risk should align with the processing purpose and its benefits, which may allow higher risk if it protects other fundamental rights.

In the proportionality evaluation one must consider the consequences and the probability of both the unwanted effects and the wanted benefits. This principle should be applied when discussing the risk of identifiability in synthetic data, while being mindful of the asymmetry of balancing individual needs against those of society.

The obligations of states to protect life and health will increase with new technological possibilities. Although it is unclear how far the obligations to fulfil the right to life extend in this context, an obligation to establish a transparent governance system can be inferred.

The European Health Data Space (EHDS) has been developed with the stated purpose of supporting societal benefits through secondary data sharing, and of improving individuals' access to and control over their personal electronic health data within the European healthcare system.[180] While the establishing of the EHDS highlights the need for sharing of health data, it is important to note that the EHDS will not solve the challenges discussed here regarding synthetic data.

The GDPR allows for the processing grounds under GDPR art. 6 and exceptions to the prohibition in art. 9 to be introduced nationally, for example for the use of synthetic data. This article does not address this in more detail beyond pointing out that this leads to variations, and that there may be an argument for exceptions in EU legislation within certain specific areas such as the public health sector.

## 6.2  Policy evaluations and the way forward

Responsible sharing to ensure data is used for the public good is an ideal ambition, reflected in EUs data strategy.[181] Ana Beduschi points to the limitations of a binary categorization of synthetic data as either personal data or not and suggests specific regulation or guidance on synthetic data generation and processing, in terms of transparency, accountability and fairness.[182]

In order to capitalize on the potential of synthetic data, it is imperative to clarify the regulations governing its use, as data sharing restrictions are currently acting as a barrier to new developments in healthcare.[183] A study conducted by the Panel for the Future of Science and Technology in the European Parliament concludes that the GDPR does not provide sufficient guidance to data controllers, and that the regulations need to be expanded and clarified.[184] The establishment of a more delineated framework for the utilization of synthetic data can facilitate advancements in AI-based medical technology. One approach to achieving this objective is to implement a distinct regulatory framework for synthetic data that sets clear specifications on the delineation of what should be considered reasonably likely identification. Alternatively, the introduction of penalties for the identification of synthetic or pseudonymized datasets could be considered. It is crucial to emphasize that such regulations must be risk-based and must consider technological developments.

EDPS warns about oligopolies and dependency of the public sector on a few powerful IT companies, so-called 'Big Tech', in their opinion on the EU Data strategy.[185] This scenario could result in the EU finding itself in a state of reliance on external entities for the advancement of AI within Europe, should the EU fail to adequately facilitate such development domestically. The present political regime in the

---

[180] Regulation (EU) 2024/2847 the European Health Data Space (EHDS).
[181] European Commission *A European strategy for data COM/2020/66 final (EU data strategy)* (2020).
[182] Ana Beduschi, 'Synthetic data protection: Towards a paradigm change in data regulation?' 11 Big Data & Society 20539517241231277, 2024.
[183] Legido-Quigley et al., 'Data sharing restrictions are hampering precision health in the European Union', 2025 (n 4).
[184] Professor Giovanni Sartorat and others, *European Parliament, The impact of the General Data Protection Regulation (GDPR) on artificial intelligence* (2020).
[185] EDPS, *Opinion 3/2020 on the European strategy for data* (2020).

### Processing of synthetic data in AI development for healthcare and the definition of personal data in EU law

USA where big technology company leaders or 'tech bros' have been granted a substantial position in the public sector- coined the 'broligarchy'- highlights the growing relevance of this concern.[186]

An important perspective that Anu Bradford, among others, has highlighted is that data protection requirements can weaken the possibilities for safeguarding other values. Although the 'Brussels effect' of EU regulatory policy is exporting norms and values from the EU to other countries, it also carries a regulatory burden. The resulting large costs are a disadvantage for smaller companies and startups, making us as a society increasingly dependent on big corporations and their owners in the future.

In the current geopolitical climate, Broomfield and Reuter express concerns not just for personal data but also regarding datasets that contain behavioural patterns that could make society vulnerable to influence.[187] They advocate restrictions in sharing all public data, including datasets that have been anonymized through generalization to not contain personal data. While the concern is valid in the current geopolitical landscape, increasing restrictions that make illegal access more difficult also have adverse effects. Restricting data processing to limit the power of big tech and their billionaire owners has inadvertently hindered the healthcare sector's ability to utilize its own collected data.[188] Hampering the development, training and validation of new AI healthcare tools in the EU market makes us more reliant on foreign big tech companies to address future challenges. Finding practical solutions that enable the use of synthetic data to develop independent tools will be an important step towards supporting the long-term goals of the EU AI Strategy while safeguarding fundamental human rights. [189]

---

[186] Lawrence Norden and Daniel I. Weiner, 'The Rise of America's Broligarchy and What to Do About It' *TIME* <https://time.com/7221154/rise-of-americas-broligarchy/> .
[187] Heather Broomfield and Lisa Marie Reutter, 'Towards a data-driven public administration: An empirical analysis of nascent phase implementation', 2021.
[188] The BigMed consortium, *Reflections on the clinical implementation of precision medicine – Experiences from BigMed, a Norwegian ICT Lighthouse project* (2021).
[189] European Commission, *COM/2018/237 Artificial Intelligence for Europe* (2018)

Processing of synthetic data in AI development for healthcare and the definition of personal data in EU law

Processing of synthetic data in AI development for healthcare and the definition of personal data in EU law

# Appendix A. Definitions and abbreviations

## Abbreviations

AEPD- Agencia Española Protección de Datos

AI Act – the EU Artificial Intelligence Act

AIA – Attribute Inference Attack

CJEU – the Court of Justice of the European Union

CNIL- The Commission Nationale de l'Informatique et des Libertés

Convention 108 – 'Convention 108 +: The Convention for the protection of individuals with regard to the processing of personal data'

DPD- 95/46/EC Data Protection Directive

EC – European Commission

ECHR- The European Convention on Human Rights

ECtHR- The European Court of Human Rights

EDPB- the European Data Protection Board

EDPS- the European Data Protection Supervisor

GDPR – the EU General Data Protection Regulation

HPC – High Performance Computing

ICESCR- International Covenant on Economic, Social and Cultural Rights

ICJ- The International Court of Justice

JS- Jensen Shannon distance

KS- the Kolmogorov Smirnov test

MIA – Membership Inference Attack

NN- The nearest neighbour analysis

Oviedo Convention- Convention on Human Rights and Biomedicine of 1996

PCJI- Permanent Court of International Justice

TEU- The Treaty on European Union

TFEU- The Treaty on the Functioning of the European Union

The Charter/ The EU Charter- The Charter of Fundamental Rights of the European Union

UDHR- The Universal Declaration of Human Rights

UN- United Nations

VCLT- Vienna Convention on the Law of Treaties

WP29- Working Party established under art. 29 of the 95/46/EC Data Protection Directive

Processing of synthetic data in AI development for healthcare and the definition of personal data in EU law

## Definitions

Real-world data- data captured based on real life events.

Anonymization - the conversion of personal data into data that cannot be used to identify any individual. This is a risk-based process, including applying both anonymization techniques and safeguards or mitigating actions to prevent identification.[190]

Health data- any data related to the physical or mental health of a natural person, especially the procurement of healthcare services revealing such information. According to Recital 35 GDPR, these include data referring to the health status of the data subject from the past, present or future. Any information on diseases, risks and disabilities- in addition to medical treatment and history- of a particular natural person therefore qualifies as health data.

Inference – 'the possibility to deduce, with significant probability, the value of an attribute from the values of a set of other attributes'[191]

Privacy risk- The technical literature usually refers to the term 'privacy risk' as a common denominator of the risk of the potential leakage of personal data from a dataset or simply identifying an individual in a dataset. The Charter of Fundamental Rights of the European Union (the EU Charter) establishes privacy and data protection as separate fundamental human rights. Art. 7 concerns respect for private and family life. Art. 8(1) covers protection of personal data. In effect, the term Privacy risk as it is used in the field of synthetic data is actually *risk of data protection breach* and usually concerns the identification of a natural person in a dataset (membership inference) or inferencing personal data related to this person (attribute inference). For the purposes of this paper, the term privacy risk is used within the context of digital personal data and the protection of this data.

Quasi-identifiers – Combinations of attributes relating to a data subject or a group of data subjects that can be used for indirect identification.

Synthetic data- data that has been artificially generated rather than captured in the real world.

---

[190] Agencia Española de Protección de Datos (AEPD) *Basic guide to anonymization by PDCP Singapore. Original title: Guía básica de anonimización Elaborada por Autoridad Nacional de Protección de Datos de Singapur (PDPC - Personal Data Protection Commission Singapore)* (2022) (n 178).
[191] WP29 *Opinion 05/2014 on Anonymisation Techniques* (2014) (n 79).

# Appendix B.     The dataset

**The original training dataset from the Florence project**

The Florence project is an interregional collaboration utilizing cancer registry data to provide clinicians in colorectal cancer with a better basis for their assessments.[192] The dataset includes 16 attributes describing the diagnosis and treatment of 30.631 colorectal cancer patients.

The data has been collected from the Norwegian Cancer Registry and has been modified initially to reduce reidentification risk. This entails removing certain parameters/attributes (columns) that are not deemed strictly necessary and removing granularity in some parameters/attributes.

The collected patients in the dataset are all colorectal cancer patients. The consequence of this is that reidentification of an individual will imply that this individual has at some stage in their life been screened for or diagnosed with colorectal cancer.  Table B- 1 summarizes the parameters/attributes of the Florence dataset.

*Table B - 1 Parameters of the Florence dataset*

| Quasi identifiers | Parameter/ attributes | Parameter code | Description | Type of value (numerical, categorical, or other) |
|---|---|---|---|---|
| | Patient serial number | P_pidKrg | A serial number for patients in the Cancer Registry (not the national ID number) | Numerical |
| | Disease incident serial number | S_pidKrg | A serial number for disease incident from the cancer registry. One patient can have several disease incidents | Numerical |
| QI | Diagnosis date | S_diagnosedato | The first date where the diagnosis is confirmed after tissue samples and/or other examinations. | Time stamp |
| | Tumor topography | S_topografIcdo3 | Location of the tumour | Categorical |
| | Tumor morphology | S_morfologiIcdo3 | Tumor genetic makeup | Categorical |
| | Surgery type | S_kir | Surgical procedure code | Categorical (binary) |
| QI | Age at the time of diagnosis | S_alderDiagnose | Patient age at the time of diagnosis | Numerical |
| QI | Gender | P_kjonn | Gender | Categorical (binary) |
| | Cause of death | P_dodsarsakkategori | Clinical codes | Categorical |
| | Underlying cause of death | P_dodsarsakUnderliggende | Clinical codes | Categorical |
| QI | Radiation | S_stralebeh | Whether patient was treated with radiation therapy | Categorical (binary) |
| QI | Person status date | P_statusdato | Date of change in person status | Time stamp |
| QI | Person status | P_personstatus | Whether the patient is deceased, lives in Norway or is lost to follow-up. | Categorical |

---

[192] Florence, 'The Florence Project home page'.

Processing of synthetic data in AI development for healthcare and the definition of personal data in EU law

| Quasi identifiers | Parameter/ attributes | Parameter code | Description | Type of value (numerical, categorical, or other) |
|---|---|---|---|---|
| QI | Radiation start date | S_datoOppstartStralebeh | Start date of radiation therapy | Time stamp |
| QI | Primary surgery date | S_datoOprPrimar | Date of primary surgery | Time stamp |
|  | Surgery priority grade | S_hastgreadKirBeh | Urgency for surgical treatment | Categorical |

### Quasi identifiers

'Quasi identifiers' or 'indirect identifiers' are attributes that could be used, either individually or in combination with other quasi-identifiers, by someone who has knowledge about that individual with the purpose of re-identifying an individual in a dataset [193]. To decide on the quasi-identifiers, it was evaluated whether this kind of information was likely to be accessible through auxiliary information, like open internet sources, media, social media or simply common knowledge. For example, the fact that someone has had surgery could be commonly known, but the clinical codes for the type of surgery will not normally be common knowledge. This leaves us with the following possible quasi-identifiers in this dataset: Date of diagnosis, Age at the time of diagnosis, Gender, Radiation, Person status date, Person status, Radiation start date and Primary surgery date.

Person status and the corresponding date describe if a person is deceased, alive and living in Norway or is lost to follow-up. The coarseness of this scale makes it a poor identifier.

It is unlikely that exact dates for treatment or diagnosis are known, as these are often the dates that the information was technically registered in the system. For example, Diagnosis date is the date of the sample or test used for diagnosis and not the date the diagnosis was actually made or communicated to the patient. One can assume that the approximate period it happened can be common knowledge, and for this case it was chosen to use the quarter it was registered as a quasi-identifier.

It is assumed that age at diagnosis and gender is known. It is deemed possible yet unlikely that all the quasi-identifiers are known for one patient and in this experiment, different combinations of this auxiliary information are tested.

The following parameters are deemed unlikely to be known by a third party: Serial numbers for Patient and Disease incident are numbers that correspond to a serial number in the cancer registry and should not be available to the attacker. Should someone gain access illegally, this would not give them more information than is in the synthetic dataset, rendering a reidentification attempt unnecessary.

---

[193] AEPD and EDPS *10 misunderstandings related to anonymisation* (2021) (n 93).

Processing of synthetic data in AI development for healthcare and the definition of personal data in EU law

Certain parameters contain specific clinical codes or information that can only be found in the patient records. The probability that anyone should have access to the actual patient records is assumed to be low, which in this dataset includes codes for Cause of death and Underlying cause of death, Surgery code, Surgery priority grade, Tumor topography and Morphology. This is why they are not included in the likely scenario.

### Sensitive attributes

The sensitive parameters in the dataset are considered to be the following:

Date of diagnosis that implies there exists a colorectal cancer diagnosis, Tumor Topography and Morphology contain clinical information regarding the tumor location and makeup, from the Surgery code one can infer clinical information, Age at diagnosis implies a diagnosis, Radiation treatment and the treatment dates contain treatment information, and the Urgency grade of surgery implies a severity of the case.

As this case is based only on EU law, an entry in the field *Cause of death* would imply the patient is deceased and the data will not fall under the material scope of the GDPR. Still, in other practical cases national laws may offer data protection to individuals also after their death, as is the case in Norway where the Florence dataset is collected.[194]

---

[194] Norwegian Ministry of Health and Care Services, *Regulation (Norway) LOV-1999-07-02-64 The Health Personnel Act (original: Helsepersonelloven)* (1999, §21 with §24. The scope of Regulation (Norway) LOV-2014-06-20-42 Processing of health data for healthcare. (Original : Pasientjournalloven) in §3 and in Regulation (Norway) LOV-2008-06-20-44 Health research (original:Helseforskningsloven) §2 also specifically state that confidentiality protection is extended after death.

# Appendix C.     Membership Inference Attack – detailed description

A Membership Inference Attack (MIA) is when an attacker attempts to determine if the data record of a specific person was included in the data used to train a machine learning model. In the context of synthetic health data, the goal of an MIA is to determine if the data from a specific patient was used to train the synthetic data generator model, by exploiting weaknesses in the data generation process. Performing a simulation of an MIA on a synthetic dataset can help investigate the robustness of the anonymization before the data is shared or made public.

## Literature review

To illustrate vulnerabilities in synthetic data generation, recent research simulates a privacy game between the data holder (challenger) and an attacker. In this game, the challenger releases a synthetic dataset, and the goal of the attacker is to determine whether a specific data record was used to derive this data. That is, the attacker wants to reveal if the synthetic release was created from a dataset containing the target record or not.[195]

Determining if the target record is part of the original dataset, underlying the synthetic release, amounts to a binary classification problem. Through a technique called shadow modelling, the attacker constructs a target record membership classifier by simulating datasets that either include or exclude the target record with machine learning.[196] Shadow modelling takes advantage of distributional differences between datasets with and without the target record to derive the classifier.

In shadow modelling, it is typically assumed that that the attacker has access to a released synthetic dataset and a target patient's data record. Some attack models further assume that auxiliary information, such as a dataset drawn from the same distribution as the training data used by the synthetic data generator, or even knowledge of the generator algorithm itself, can be exploited. However, these assumptions are subject to criticism for being unrealistic in many practical contexts.

A data holder aiming to avoid sensitive information disclosure through a synthetic data release may take steps to remove potential links to auxiliary information. This is particularly true in the case of synthetic healthcare data, where it is highly unlikely that an attacker would have access to data from the same distribution as the original dataset used to train the generator. Furthermore, the technical complexity of configuring attack algorithms, along with a substantial demand for computational power and time, may further limit the feasibility of such attacks.

---

[195] Gadotti et al., 'Anonymization: The imperfect science of using data while preserving privacy', 2024 (n 44)44.
[196] Florent Guépin et al., 'Synthetic is all you need: removing the auxiliary data assumption for membership inference attacks against synthetic data' (European Symposium on Research in Computer Security, 2023).

Several studies indicate that MIAs on synthetic data can indeed be computationally intensive.[197] For example, the DOMIAS framework leverages density estimators to target overfitting in the synthetic data generator and requires significantly more computational resources than simpler inference techniques. Another study uses techniques from contrastive representation learning to model subtle differences between original and synthetic distributions, which illustrates that the effectiveness of MIAs depends on data availability and the attacker's computational resources, as they require creating numerous variations and comparing them across models.[198]

## Method

To test the legal assessments discussed in this paper, an MIA case study was conducted on a synthetic dataset, using data samples from a research project named Florence. Although a single simulation study may not provide generalizable conclusions for other contexts, it underscores key success factors and limitations of this type of attack. Challenges in simulating an attack may render its practical implementation infeasible, indicating a reduced risk of sensitive information disclosure through MIAs.

### Attack simulator

The attack simulator is based on a privacy game between a data holder and an attacker. The game involves several rounds, each starting with a randomly chosen target record from the original dataset. In each round, the data holder releases ten datasets: five including the selected target record and five excluding it. This setup provides the data holder with a ground truth for each released dataset, allowing an evaluation of the attacker's success in determining target record membership.

To introduce variability among the released data and to enable an analysis of the attacker's success, each dataset release is derived from a random sample of the original data. As prior research suggests that not all records are equally vulnerable, a different target record was randomly selected in each run.[199]

---

[197] Guépin et al., 'Synthetic is all you need: removing the auxiliary data assumption for membership inference attacks against synthetic data' (2023) ; Boris van Breugel et al., *Membership Inference Attacks against Synthetic Data through Overfitting Detection* (PMLR, 2023) ; Ziqi Zhang, Chao Yan and Bradley A Malin, 'Membership Inference Attacks against synthetic health data' 125 Journal of biomedical informatics 103977, 2022

[198] Dong-Yu She and Kun Xu, 'Contrastive self-supervised representation learning using synthetic data' 18 International Journal of Automation and Computing 556, 2021

[199] Meeus et al., 'Achilles' heels: vulnerable record identification in synthetic data publishing' (2023) (n 48).

*The simulation algorithm*

The attack model assumes that the presence or absence of a target record will introduce shifts in the data distributions that can be detected by the attacker. In the simulation, the attacker attempts to determine whether the target record of a specific patient contributed to creating the released data. The simulation builds on a state-of-the-art MIA technique called shadow modelling, where the attacker trains a target record membership classifier using auxiliary data similar to the released data. [200] The MIA is contextualized in a practically likely scenario, restricting the attacker to access only the released dataset and the target record.

To simulate an attack, the approach is based on *Guépin and others*.[201] Here, the attacker circumvents the lack of auxiliary information for shadow modelling by creating a synthetic version of the dataset released to the attacker, referred to as the synthetic auxiliary. This synthetic auxiliary data is used to construct a shadow dataset for training a target record membership classifier.

In Guépin and others,[202] the shadow dataset is created via three steps: 1) sampling a subset from the synthetic auxiliary, 2) generating a synthetic version of the auxiliary subsample by fitting a synthetic data generator and 3) extracting information about the synthetic auxiliary subsample into a feature vector. These three steps produce only a single record for the shadow dataset. Hence, creating a shadow dataset of N samples requires fitting N generators in step 2, inducing a significant computational overhead for the attack.

*Feasibility adjustments to the simulation algorithm*

Implementing the algorithm described in *Guépin and others* on the HPC system exceeded the wall-clock time limit of one week, which has been imposed for security reasons.[203] To improve feasibility, step 2 of the simulation algorithm was modified. Instead of training N generators, only two were trained: one generator fitted to the synthetic sub-sample combined with the target record, and another generator on the sub-sample without the target record. These two generators are used to create data samples for feature extraction. This modification allowed us to reduce the computational load while maintaining the integrity of the simulation process.

---

[200] Meeus et al., 'Achilles' heels: vulnerable record identification in synthetic data publishing' (2023) (n 48).
[201] Guépin et al., 'Synthetic is all you need: removing the auxiliary data assumption for membership inference attacks against synthetic data' (2023) (n 197).
[202] Guépin et al., 'Synthetic is all you need: removing the auxiliary data assumption for membership inference attacks against synthetic data' (2023) (n 197), also used in Meeus et al., 'Achilles' heels: vulnerable record identification in synthetic data publishing' (2023) (n 48). Code repository available here: Meeus et al., 'Code repository- MIA synthetic' <https://github.com/computationalprivacy/MIA-synthetic> accessed 17.3.2025
[203] Guépin et al., 'Synthetic is all you need: removing the auxiliary data assumption for membership inference attacks against synthetic data' (2023) (n 197).

Processing of synthetic data in AI development for healthcare and the definition of personal data in EU law

The features extracted from the synthetic auxiliary subsample include marginal descriptive statistics, frequency counts of each data attribute and the pairwise Spearman correlation, similar to *Stadler and others.*[204] Features that can be estimated without knowing the target record were selected, since this information may in practice be uncertain to the attacker.

*Performance evaluation*

To estimate the target record membership, the attacker applies the trained membership classifier to the released dataset. The success of the attack is determined by comparing the attacker's membership estimate to the data holder's ground truth. A performance score, such as the ROC AUC, is used to summarize the attacker's performance. The ROC AUC ranges from zero to one. A score close to one suggests a highly successful attack, while a score close to 0.5 reflects that the attacker is making random guesses about the target record membership. If the score is below 0.5, it indicates that the attack model is flawed since the attacker is consistently making incorrect decisions about record membership.

## Experimental setup

As is common in MIA simulations, it is assumed the attacker has access to the synthetic dataset, information about the original target record, and the details of the generator that was used to create the synthetic release.

*The original data*

In each round of the attack simulation, a dataset of 1000 records were sampled from the Florence data as the original private data in this case study.

To start, the vulnerability of the original data was investigated to assess the potential difficulty of attacks. A cluster analysis was performed by grouping together similar data records. Datasets with highly distinguishable records may present greater vulnerability to attacks, whereas more homogenous datasets are less susceptible. Using K-means clustering, each data sample is segmented into a maximum of 15 clusters, and the minimum number of records within any cluster were measured. Larger clusters suggest greater similarity among records, in which case such attacks may be less effective.

---

[204] Theresa Stadler, Bristena Oprisanu and Carmela Troncoso, 'Synthetic data–anonymisation groundhog day' (31st USENIX Security Symposium (USENIX Security 22), 2022)).

Processing of synthetic data in AI development for healthcare and the definition of personal data in EU law

*The synthetic data generator*

The selected generator algorithm is based on a publicly available implementation of the conditional variational autoencoder network (TVAE), designed to handle tabular data.[205] TVAE was chosen as they may be considered less computationally demanding than Generative Adversarial Networks (GANs), although this can vary based on specific implementations and tuning.[206] The target record membership classifier used for shadow modelling is based on a publicly available implementation of the random forest model.[207]

*Synthetic data assessment*

The goal in synthetic data generation is to have data mirroring key characteristics of the original dataset, while avoiding direct replication of its original records. As the attack model relies on descriptors of the data attribute distributions to detect target record membership, a key element of the synthetic data is to reflect the distributional information of the original dataset.

To evaluate the ability of the TVAE to generate synthetic data preserving the distributional characteristics of the original data, a preliminary study was conducted to assess the synthetic outputs. In this study, TVAEs were fitted to ten random subsamples drawn from the Florence data and generated corresponding synthetic datasets to each subsample. The synthetic and the original subsamples were compared through quantitative similarity measures. These measures include a nearest neighbour analysis of information replication, a distance metric and statistical hypothesis testing of the alignment of data distributions.

*Nearest neighbour analysis of record similarity*

The nearest neighbour (NN) analysis measures the average distance from each synthetic record to its closest counterpart in the original data. While detecting copies of original records in synthetic data is often analysed through pairwise comparisons, the NN method extends beyond considering only corresponding record pairs. The NN score ranges from zero to one, representing the average scaled distances, where smaller values indicate more resemblance between original and synthetic records.

---

[205] van der Schaar labs, 'SynthCity- plugins repository ' <https://github.com/vanderschaarlab/synthcity/tree/main/src/synthcity/plugins> accessed 17.03.2025.
[206] Lei Xu et al., 'Modeling tabular data using conditional gan' 32 Advances in neural information processing systems, 2019.
[207] Tin Kam Ho, *Random decision forests* (IEEE, 1995) , repository: 'Scikit learn- RandomForestClassifier v. 1.6.1' <https://scikit-learn.org/stable/modules/generated/sklearn.ensemble.RandomForestClassifier.html> accessed 17.03.2025.

### Analysis of distributional similarity

The distribution of data attributes between the original and the synthetic data, three statistical measures were considered based on the Jensen Shannon (JS) distance, the Kolmogorov Smirnov (KS) test and Chi squared test. Each measure provides a score from zero to one for the resemblance between the original and synthetic distributions.

A JS distance score of zero represents two identical distributions, while a score of one represents maximum dissimilarity. The distance between each pair of corresponding attributes in the synthetic and original data were calculated and the average distributional similarity across the dataset was reported. This measure is applicable to both categorical and continuous attributes.

The KS test is designed primarily for continuous data. For totally different distributions, the test score is zero, while the score is closer to one for identical distributions. The score derives from the two-sample Kolmogorov-Smirnov test statistic, averaged across the continuous attributes in the data.

For categorical data, the Chi squared test is used. The score is based on the test statistic, reflecting the similarity between observed and expected frequencies in attribute distributions. The Chi squared score is averaged across all categorical attributes, and its interpretation aligns with the KS test, with higher values indicating greater similarity.

### *Attack settings*

To assess the vulnerability of a synthetic release and the efficiency of an attack model, four attack settings are explored, differing in the data that is used to represent the synthetic release.

*Original data testing*

In this setting, the synthetic data given to the attacker is actually the original data, illustrating an upper bound on the expected performance of the attack. When the released dataset has a perfect resemblance to the original data, one can expect the attacker to be successful given an adequate attack model.

*Upper bounds scenario*

A fully synthetic dataset based on all the variables in the original dataset is given to the attacker. The synthetic release is created by fitting a synthetic data generator to the original dataset and sampling records from this generator. In this scenario, the attacker carries out an MIA using data with the complete set of variables.

Processing of synthetic data in AI development for healthcare and the definition of personal data in EU law

*Likely scenario*

Compared to the fully synthetic scenario, in this case the released synthetic data contains only a subset of the columns in the original data (selected quasi-identifiers). The success of the attack will increase with a lower number of entries in a dataset and with a larger proportion of anomalies, as you may find in datasets for rare disease or small population studies.[208] The expectation is therefore that an attack against the synthetic quasi will be more difficult due to fewer columns than the fully synthetic.

### Computational resources

Per attack setting, 50 rounds of attack simulations were executed in total. Using the feasibility adjusted simulation algorithm, each round could take up to two hours. The simulations were carried out on the Colossus high-performance computing (HPC) cluster hosted by the University of Oslo. Through the Colossus HPC system, 300 GB of memory and 64 cores was requested to parallelize the rounds of the attack simulation.

### Code availability

The source code for the attack simulation will be released together with the publication of this manuscript. [209]

## Results

The goal in simulating this case study is to illustrate the potential attack outcomes and challenges and practical limitations related to conducting an MIA on randomly selected target records.

Through a K-means clustering analysis, it was identified that the smallest clusters in the original dataset contained only around five percent of the records. Smaller clusters indicate more diversity among records in the dataset, in which case MIA can be more effective if the target record belongs to any of the smaller clusters.

### Synthetic data assessment

Through a pre-study, the ability of TVAE to produce synthetic data capturing the distributional information from the original data without fully replicating the original information was evaluated. In a pairwise comparison between corresponding synthetic and original records no pair of rows were identical, verifying that none of the original records were reproduced in the synthetic dataset.

---

[208] Shalini Dwivedi, 'Synthetic data and European General Data Protection Regulation: Ethics, quality and legality of data sharing' 6 Journal of Data Protection & Privacy 332, 2024.
[209] *To be included upon publication.*

Additionally, a nearest neighbour record similarity analysis was conducted with estimated statistical descriptors of distributional similarity, which are presented in Table C- 1.

A key quality of the synthetic data is to reflect the distributional information of the original dataset, since the attack model assumes distributional shifts in the attributes will occur depending on presence of a target record. Recall that the JS distance indicates the average similarity between the distributions of the categorical and continuous attributes in the dataset, while the KS test is mainly for continuous data and the Chi squared test is for categorical data. In terms of the JS score, more similar distributions have a score closer to zero, while the KS and Chi scores are close to one. The nearest neighbour analysis gives the average Euclidean distance between synthetic and original records, scaled to a range from zero to one.

*Table C- 1 Quantitative measures for comparing synthetic and original datasets*

| Nearest neighbour distance | Jensen Shannon Distance | Kolmogorov Smirnov Test | Chi Squared Test |
| --- | --- | --- | --- |
| 0.34 +/- 0.02 | 0.011 +/- 0.001 | 0.83 +/- 0.01 | 0.634 +/- 0.06 |

The nearest neighbour distance in Table C- 1 suggests a notable degree of similarity between any pair of synthetic and original records. Furthermore, the JS score being closer to zero indicates a good overall approximation to distribution of original data. Examining attribute distribution similarity separately for continuous and categorical variables, the KS test shows that the distribution of continuous attributes is more similar that that of the categorical attributes. In summary, Table C- 1 shows that the TVAE is capable of generating synthetic data that captures the distributional information of the original dataset, without replicating the original records.

### Attack outcomes

Figure C- 1 shows a swarm plot of the attacker's success in four different settings: 1) A fully synthetic dataset including all variables in the original data (Upper bounds scenario); 2) A synthetic dataset including only the quasi-identifiers (Likely scenario); 3) Testing the MIA methodology on the original data being rather than the synthetic dataset, including all variables (Original data). Each data point in the plot depicts the ROC AUC score for an attack against a specific target record.

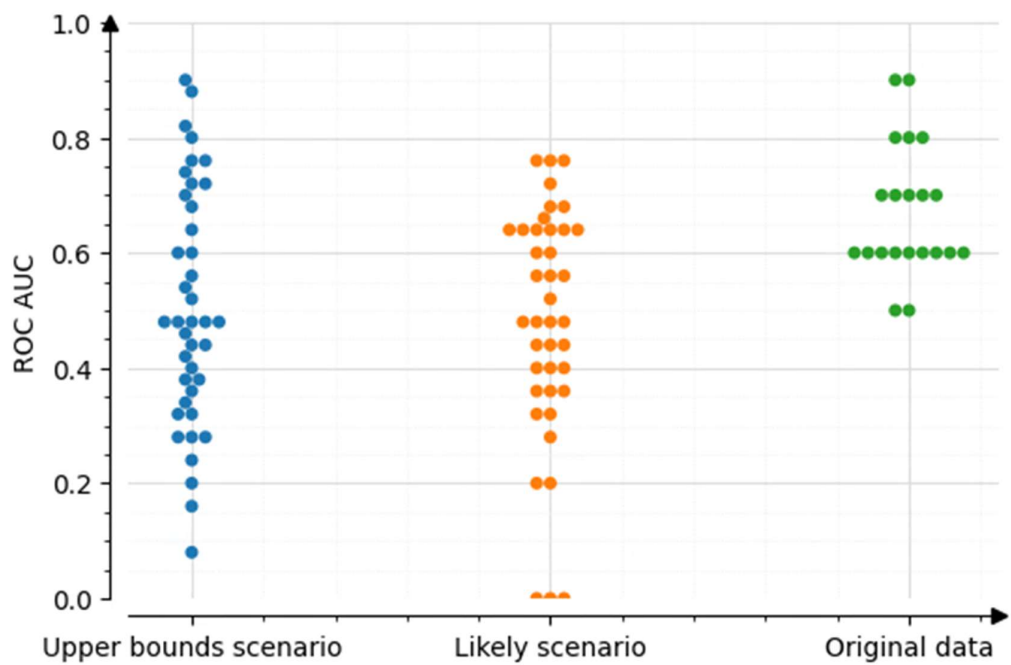

*Figure C- 1 Swarm plot of the attack simulation results*

The swarm plot shows that the attacker can achieve considerable success on the fully synthetic and quasi synthetic datasets for certain target records (ROC AUC above 0.7). From the attacker's perspective, this outcome indicates that target record membership may be successfully inferred, and that the synthetic data release can be vulnerable to sensitive information disclosure.

From the data-holder's perspective, releasing less detailed information proves somewhat more resilient to attacks, as indicated by the slightly lower ROC AUC scores on the synthetic quasi data compared to the fully synthetic release. The AUC scores between 0.5 and 0.7 on the quasi and fully synthetic releases suggests that the attacker faces more difficulty in accurately detecting the presence of the target record. This is especially true for target records that resemble other records in the dataset, as less unique records provide weaker signals for identification. Therefore, removing outliers and ensuring greater homogeneity in the released data will reduce its susceptibility to MIAs.

While other studies leverage extensive auxiliary information in shadow modelling, this study considers a potentially more realistic scenario in which the attacker does not have access to the entire patient record. [210] This assumption created a more challenging situation for the attacker. Where the swarm plot shows AUC scores below 0.5, it indicates that the attacker consistently fails to classify the target records, suggesting possible issues with the attack model. Even when attacking the original data with access to authentic information, the attacker struggles to detect the target record. This indicates that the attack model may be inadequate for fully capturing the privacy risks associated with synthetic data release.

---

[210] Gadotti et al., 'Anonymization: The imperfect science of using data while preserving privacy', 2024 (n 44)44.

**Implementation challenges**

As shown in Figure C- 1, AUC scores below 0.5 indicate significant issues with the target record membership classifier failing to identify target record membership. Several factors can contribute to this underperformance. In these simulations, two main factors were identified, namely sample noise and suboptimal feature extraction methods used to construct the shadow data for training the classifier. These challenges highlight the importance of carefully configuring the attack to maximize the changes of success.

In this case study, the attacker is only given the released dataset to infer target record membership, without any additional information typically assumed in MIA models.[211] To implement shadow modelling, the attacker creates additional synthetic datasets from the release and uses these to train the target record classifier. However, using an imperfect generator to create synthetic data from other synthetic data can amplify noise and introduce a distributional shift between the shadow data and the released data. This shift may render the classifier trained on the shadow data inappropriate for the released data due to differences between the datasets.

Another factor affecting classifier performance is suboptimal feature extraction. The success of the attack depends on whether the presence of the target record has a detectable impact on any of the features, and whether the synthetic dataset has preserved these features from the original data. Thus, a difficulty related to shadow modelling is defining features that are sensitive to the influence of a single target record on the data distribution. The features used to train the classifier are derived from descriptive statistics, count statistics and the pairwise attribute correlations. Although a thorough review and improvement of feature extraction methods is beyond the scope of this article, fitting a classifier without sufficient features may lead to overfitting and reduced generalization abilities.

# Conclusion

In summary, this case-study demonstrates that synthetic data may offer some degree of protection against MIA, with attack effectiveness strongly influenced by the choice of target record. Notably, the results suggest that the attack model may not be adequate to fully capture the risk associated with synthetic data release. However, comprehensive reassessment and refinement of the attack model is beyond the scope of this paper and is reserved for future work. This contribution highlights the need for a more nuanced understanding of the vulnerabilities inherent in synthetic data and encourages further research to enhance and refine attack methodologies in this domain.

---

[211] Gadotti et al., 'Anonymization: The imperfect science of using data while preserving privacy', 2024 (n 44)44.

History may have taught us to anticipate continued development in both privacy-preservation techniques and attack methods. As open-source code is continuously being made available, attackers have the opportunity to adapt and apply the most recent attack strategies. Factors influencing the outcome of the attack include the information available to the attacker, the quality of the synthetic data and the adequacy of the attack methodology. It is commonly assumed that the attacker knows the target record, the synthetic data generator algorithm and has access to some information of the original data distribution. However, these assumptions may in certain contexts be unrealistic and limit the generalizability of MIA success.

Moreover, even if the attacker is successful, the implications depend on the nature of the data and the sensitivity of the information revealed. A successful attack provides only an estimate of target record membership, which may be difficult to affirm unless the data holder is confronted and reveals the ground truth.